\documentclass[prb,twocolumn,amsmath,amssymb,floatfix,superscriptaddress,nofootinbib]{revtex4-2}
\usepackage{graphicx}
\usepackage{bm}
\usepackage{booktabs}
\usepackage{hyperref}
\usepackage{amsthm}

\graphicspath{{figures/}}
\newcommand{\Am}{\bm{A}}
\newcommand{\bR}{\mathbf{R}}
\newcommand{\bG}{\mathbf{G}}
\newcommand{\bk}{\mathbf{k}}

\begin{document}

\title{Dipolar order across Bravais lattice space:\\
classification, spin waves, and a four-attractor phase diagram}

\author{J. Batle}
\affiliation{CRISP -- Centre de Recerca Independent de sa Pobla, sa Pobla, Balearic Islands, Spain}
\affiliation{Departament de F\'{\i}sica, Universitat de les Illes Balears,
07122 Palma de Mallorca, Balearic Islands, Spain}

\begin{abstract}
Every crystal is a Bravais lattice decorated by a basis, so the dipolar ordering of the
fourteen Bravais lattices is the natural starting point for any systematic theory of
dipolar magnetism in three dimensions. We determine it here in a single common framework:
the interaction tensor is Ewald-summed, the classical ground state is obtained by
minimising the lowest Luttinger--Tisza band over the entire Brillouin zone, and the linear
spin-wave spectrum with its zero-point corrections is computed for every lattice whose
order is collinear. Three results emerge that are not properties of individual lattices but
of the landscape. First, two structural principles --- the exact tracelessness of the
dipolar tensor in three dimensions, and its identical vanishing at $\mathbf k=0$ for every
cubic-symmetric lattice --- explain the ordering type, the absence of first-order cubic
anisotropy, and the systematics of the zero-point moment reduction. Second, exactly one
lattice defeats the Luttinger--Tisza construction: for face-centred orthorhombic the
optimal eigenvector is not circular, the single-$\mathbf k$ state is a spin-density wave
of non-constant length, and the tabulated energy is a strict lower bound; direct supercell
minimisation gives the true ground state. Third, optimising each family over its free
metric parameters collapses the whole of Bravais space onto only \emph{four} attractors,
with body-centred tetragonal the global optimum and a single interior optimum at
rhombohedral $\alpha=62.42^\circ$ lying below face-centred cubic. Three independent
published benchmarks are reproduced.
\end{abstract}

\maketitle

\section{Introduction}

Magnetic dipoles on a lattice interact through a force that is long-ranged, anisotropic, and
of a sign that depends on the geometry of the bond: two moments attract when placed head to
tail and repel when placed side by side. A crystal of such moments must therefore reconcile
two things at once --- where the sites sit, and which way the moments point --- and the two
are not independent. This makes the dipolar lattice problem qualitatively different from the
central-force problems of classical lattice-energy theory, where only the positions matter,
and it is the reason that quite simple questions about it have remained open.

The problem is not new. Luttinger and Tisza~\cite{LuttingerTisza1946,LuttingerTisza1947}
introduced the method still used to solve it and applied it to the cubic lattices, finding
that simple cubic orders antiferromagnetically while the body- and face-centred lattices order
ferromagnetically. The chain-forming tendency of dipolar matter was analysed in the
electrorheological literature by Halsey and Toor~\cite{HalseyToor1990} and Tao and
Sun~\cite{TaoSun1991}, who identified the body-centred tetragonal structure at
$c/a=\sqrt{2/3}$; Gross and Wei~\cite{GrossWei2000} obtained its energy to machine precision
by annealing in the space of chain arrangements, and Groh and Dietrich~\cite{GrohDietrich2001}
recovered the same structure within a Stockmayer phase diagram. Dipolar interactions on
frustrated lattices have their own extensive literature, the spin-ice
compounds~\cite{BramwellGingras2001,denHertogGingras2000,Melko2001} being the best known case,
and two-dimensional dipolar arrays have been surveyed
separately~\cite{PolitiPini2002,CzechVillain1989,Vedmedenko1998}.

What is missing from this record is the landscape. The results above were obtained for
particular structures, by different methods, in different normalisations, and with different
conventions for handling the conditional convergence of the dipolar sum --- a matter that
affects any state carrying a net moment. There is no systematic account of what the dipolar
interaction does across the whole of translational symmetry, and consequently no way to tell
which features of the known results are general and which are peculiar to the cubic cases
that happen to have been studied.

This paper supplies that account. We treat all fourteen Bravais lattices in a single
framework: the interaction tensor is Ewald-summed in a stated convention, the classical
ground state is obtained by minimising the lowest Luttinger--Tisza band over the entire
Brillouin zone rather than assuming a uniform state, and the linear spin-wave spectrum with
its zero-point corrections is computed for every lattice whose order is collinear. We then
optimise the free metric parameters within each family, which turns the catalogue into a
phase diagram. Three independently published values are reproduced as validation.

A reader should take three things from what follows. The first is that the ordering type is
governed by the centring of the lattice, and that this rule --- stated by Luttinger and Tisza
for the cubic cases --- holds across the whole family for a reason that is geometric and can
be stated in a sentence. The second is that the quantum corrections do not measure what one
might expect: the zero-point moment reduction sorts the lattices by anisotropy rather than by
ordering type, so that the most symmetric lattices have the largest corrections. The third is
that the energetic optimum is not the densest packing, in contrast to the two-dimensional
case, and that exactly one of the fourteen lattices defeats the Luttinger--Tisza construction
outright --- a failure that is usually treated as a formal caveat and here is a real one.

\subsection{Context: why dipolar lattice sums remain a live problem}

The dipolar interaction is peculiar among pair interactions in three ways, and all three
are responsible for the difficulty of the problem treated here.

It is \emph{anisotropic}. Unlike the central-force potentials of classical lattice-energy
theory --- Lennard-Jones, Morse, or the inverse powers whose lattice minimisation is the
subject of the Epstein zeta literature~\cite{Rankin1953,Cassels1959,Ennola1964,Diananda1964,%
Montgomery1988,SarnakStrombergsson2006} and of the universal-optimality
programme~\cite{CohnKumar2007,Cohn2022E8Leech,BetermanZhang2015,Betermin2019Morse,%
Betermin2021computerassisted} --- the dipolar kernel depends on the orientation of the bond
relative to the moments. Each site therefore carries an orientational degree of freedom
that must be optimised jointly with the geometry, and the scalar machinery does not apply.

It is \emph{long-ranged and only conditionally convergent}. The lattice sum of $r^{-3}$
kernels in three dimensions diverges logarithmically in absolute value, so its value depends
on the order of summation, or equivalently on the sample shape. Ewald's
construction~\cite{Ewald1921}, in the form developed for lattice sums by Nijboer and de
Wette~\cite{NijboerdeWette1957} and by Smith and Ashcroft~\cite{SmithAshcroft1988},
resolves this into two absolutely convergent pieces at the cost of fixing a convention,
which must then be stated explicitly and held fixed across any comparison.

It is \emph{frustrating in the technical sense}. The competition between the attractive
head-to-tail and repulsive side-by-side configurations means that no local rule determines
the global optimum, and highly degenerate manifolds are common. This is the mechanism
behind spin ice~\cite{BramwellGingras2001,denHertogGingras2000,Melko2001,MelkoGingras2004,%
Isakov2005}, where long-range dipolar interactions acting on a pyrochlore lattice produce
an extensively degenerate low-energy manifold, and it is why simple cubic --- the most
symmetric lattice in the table --- has the largest zero-point moment reduction of all.

The method used throughout is due to Luttinger and Tisza~\cite{LuttingerTisza1946,%
LuttingerTisza1947}, who replaced the hard constraint $|\mathbf S_\bR|=1$ by the single
weaker condition $\sum_\bR|\mathbf S_\bR|^2=N$, reducing the problem to an eigenvalue
problem at each wavevector. The construction was extended and formalised by Lyons and
Kaplan~\cite{LyonsKaplan1960}, by Litvin~\cite{Litvin1974}, and by Friedman and
Felsteiner~\cite{FriedmanFelsteiner1974}; the circumstances under which the relaxed
solution fails to satisfy the original constraint have been discussed since the
method's inception, and one instance of that failure is documented in
Section~\ref{sec:oF} below. Applications to dipolar crystals specifically include the
Ising-dipolar work of Fern\'andez and Alonso~\cite{FernandezAlonso2000}, the
electrorheological-fluid literature initiated by Halsey and Toor~\cite{HalseyToor1990} and
Tao and Sun~\cite{TaoSun1991,TaoSun1991b}, the ground-state searches of Gross and
Wei~\cite{GrossWei2000} and of Groh and Dietrich~\cite{GrohDietrich2001}, the simulation
studies of Weis and Levesque~\cite{WeisLevesque1993} and Wei and Patey~\cite{WeiPatey1992},
and the systematic tabulation of dipolar lattice sums by Johnston~\cite{Johnston2016}.
Two-dimensional dipolar arrays have their own substantial
literature~\cite{PolitiPini2002,CzechVillain1989,Vedmedenko1998}, and the planar
counterpart of the present survey is reported
separately~\cite{BatleCiftja2020,Batle2021}.

Experimentally, dipolar-dominated magnetism is realised in rare-earth
insulators, in molecular magnets crystallising on low-symmetry
lattices~\cite{FernandezAlonso2000}, in ferrofluids and
electro/magnetorheological suspensions~\cite{Rosensweig1985,ChenZitterTao1992,%
Dassanayake2000,Martin1998}, and in lithographically patterned nanomagnet arrays, where
the lattice geometry is a fabrication choice rather than a property of a compound. In the
last of these settings, knowing which geometry is energetically optimal is a design
question rather than an academic one.

\subsection{Scope}

Throughout, one dipole per primitive cell and classical moments of fixed length free to
choose their direction. The Luttinger--Tisza analysis is single-$\bk$ by construction; where
that description fails, as it does for one lattice, we abandon it and minimise directly over
unit moments in a supercell, a procedure that admits multi-$\bk$ and non-collinear states; lattices with a basis are a separate problem, in
which optical magnon branches and frustration within the basis introduce phenomena that
have no counterpart here.

\section{Formulation}

\subsection{The functional}
\begin{figure}[t]
\centering
\includegraphics[width=\columnwidth]{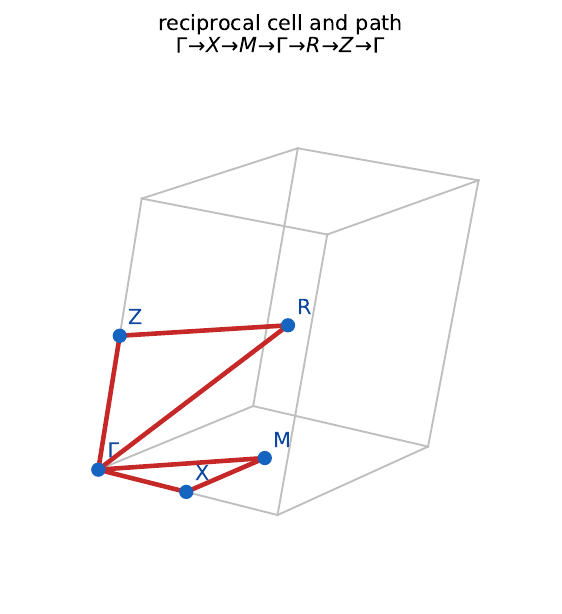}
\caption{The reciprocal cell and the band-structure path used for every lattice. A generic
triclinic cell admits no standard high-symmetry labelling, so the corners of the reciprocal
parallelepiped are used uniformly; for the high-symmetry members the path coincides with
the conventional one.}
\label{fig:bz}
\end{figure}

For a lattice $\Lambda$ and a modulated configuration
$\mathbf S_\bR=\mathrm{Re}(\mathbf m_\bk e^{i\bk\cdot\bR})$ the energy per site is
$\tfrac12\mathbf m_\bk^{\dagger}\Am(\bk)\mathbf m_\bk$ with
\begin{equation}\begin{split}
\Am(\bk;\Lambda)=\sum_{\bR\in\Lambda\setminus\{0\}}
\frac{I_3-3\hat\bR\hat\bR^{T}}{|\bR|^{3}}\,e^{i\bk\cdot\bR},
\\[2pt]
e_{\min}(\Lambda)=\min_{\bk\in\mathrm{BZ}}\tfrac12\lambda_{\min}\bigl[\Am(\bk)\bigr].
\end{split}\label{eq:functional}
\end{equation}
The sum is only conditionally convergent and is evaluated by Ewald
summation~\cite{Ewald1921,NijboerdeWette1957,BornHuang1954,AshcroftMermin1976}. Band structures are reported along the path of Fig.~\ref{fig:bz} throughout. We fix the
scale by the shortest lattice vector, $\min_{\bR\neq0}|\bR|=1$, so energies are in units
of $\mu^{2}/d_{nn}^{3}$, and we use the tinfoil (needle-shaped-sample) convention
throughout: the $\bG=0$ term is dropped at $\bk=0$. This matters only for states with a
net moment, and it must be fixed before any comparison across the table or with the
literature is meaningful.

\subsection{Two properties of the dipolar tensor}\label{sec:principles}
Two elementary properties of the interaction tensor are used repeatedly below. Between them
they account for the ordering type of each lattice, for the absence of first-order cubic
anisotropy, and for the systematics of the quantum corrections.

\paragraph{Tracelessness of the dipolar tensor.}\label{prop:trace}
$\operatorname{tr}(I_3-3\hat\bR\hat\bR^{T})=0$ for every $\bR$, hence
$\operatorname{tr}\Am(\bk)=0$ for every lattice and every $\bk$.

In two dimensions the corresponding trace is $-1$, and the resulting scalar
Epstein-zeta shadow organises the planar theory. In three dimensions it vanishes: the
eigenvalues of $\Am$ always sum to zero, so $\lambda_{\min}\le0$ always, with equality only
if $\Am\equiv0$.

\paragraph{Vanishing at $\bk=0$ for cubic symmetry.}\label{prop:cubic}
If $\Lambda$ has cubic point symmetry then $\Am(0)=\alpha I_3$; combined with
Proposition~\ref{prop:trace}, $\alpha=0$ and $\Am(0)\equiv0$.

Numerically the three cubic lattices give an anisotropy spread
$\lambda_{\max}-\lambda_{\min}$ of $8.9\times10^{-16}$ (sc), $5.3\times10^{-15}$ (bcc) and
$0$ (fcc), confirming Proposition~\ref{prop:cubic} to machine precision. The consequence
is important for reading the table: \emph{for a cubic lattice the ferromagnetic easy axis
is completely degenerate at this order}. The cubic easy directions reported in the
literature --- cube edges for simple cubic, body diagonals for bcc and fcc~\cite{Syro2006}
--- arise from $1/S$ corrections beyond the linear theory used here, and any easy axis our
diagonalisation returns for a cubic lattice is numerical noise. Non-cubic lattices are
unaffected: body-centred tetragonal, for example, has eigenvalues
$(-6.100200,-5.327481,-5.327481)$, a genuine anisotropy of $0.773$ with a unique easy axis.

\subsection{Spin waves and zero-point corrections}
Where the ground state is ferromagnetic, $\bk_0=\Gamma$ with easy axis $\hat n$, we choose a
local frame $(\hat u,\hat v,\hat n)$, write $E_{nn}=\hat n\cdot\Am(0)\hat n$, and set
\begin{equation}\begin{split}
P(\bk)=\hat u\Am(\bk)\hat u-E_{nn},\quad
Q(\bk)=\hat v\Am(\bk)\hat v-E_{nn},
\\[2pt]
R(\bk)=\hat u\Am(\bk)\hat v .
\end{split}\end{equation}
The Holstein--Primakoff expansion~\cite{HolsteinPrimakoff1940} followed by a Bogoliubov
rotation~\cite{Bogoliubov1947} gives
\begin{equation}
\varepsilon(\bk)=\sqrt{P Q-R^{2}},\qquad
A_\bk=\tfrac12(P+Q),
\label{eq:magnon}
\end{equation}
and the standard zero-point quantities
\begin{equation}
\delta S=\Bigl\langle \frac{A_\bk}{2\varepsilon(\bk)}-\frac12\Bigr\rangle_{\mathrm{BZ}},
\qquad
e_{\mathrm{qu}}=e_{\min}+\tfrac12\bigl\langle \varepsilon(\bk)-A_\bk\bigr\rangle_{\mathrm{BZ}},
\label{eq:zeropoint}
\end{equation}
$\delta S$ being the zero-point reduction of the ordered moment and $e_{\mathrm{qu}}$ the
magnon-corrected energy; the general definition of $\delta S$ as
$S-\langle S^{z}\rangle=\langle a^{\dagger}a\rangle$ averaged over the zone is standard.
The requirement $\varepsilon(\bk)\ge0$ throughout is the dynamical-stability test: an
imaginary mode signals that the assumed state is not a true minimum.

\paragraph{The discontinuity at $\Gamma$.}\label{rem:gammajump}
The magnon dispersions show an apparent jump where the path passes through $\Gamma$. This
is not a plotting artefact. The dipolar $\Am(\bk)$ is non-analytic at $\bk=0$: the $\bG=0$
term of the reciprocal sum contributes $(4\pi/V)\,\hat\bk\hat\bk^{T}$, whose limit depends
on the direction of approach. Consequently $\varepsilon(\bk\to0)$ is direction-dependent,
and since the path $\Gamma\to X\to M\to\Gamma\to R\to Z\to\Gamma$ visits $\Gamma$ three
times --- arriving along one direction and leaving along another --- a genuine
discontinuity appears there. For body-centred tetragonal, for instance, the limits are
$1.958$ approaching along $\overline{M\Gamma}$, $3.680$ along $\overline{\Gamma R}$ and
$0.773$ along $\overline{Z\Gamma}$. This is the long-wavelength magnetostatic regime
familiar from dipolar magnonics, in which the mode frequency depends on the propagation
direction relative to the magnetisation; it is a physical feature of dipolar spin waves and
not a numerical one.

A finite ordering wavevector does not by itself require the rotating-frame machinery, and
it is worth being precise about which cases do. When every fractional component of $\bk_0$
lies in $\{0,\tfrac12\}$ one has $\cos(\bk_0\cdot\bR)=\pm1$, so neighbouring moments are
simply parallel or antiparallel: the state is a \emph{collinear} multi-sublattice
antiferromagnet or stripe, and conventional two- or four-sublattice linear spin-wave
theory applies, exactly as for the square-lattice Heisenberg antiferromagnet. Only genuinely
non-collinear (incommensurate) order, where the moment direction rotates continuously from
cell to cell, requires a local-axis expansion. Of the eight finite-$\bk$ entries in
Table~\ref{tab:main}, seven are commensurate and hence collinear; only face-centred
orthorhombic, at $\bk_0\simeq(0.15,0.15,0.30)$, is truly incommensurate.

Both cases are treated here: the ferromagnets by Eq.~\eqref{eq:magnon} and the seven
commensurate antiferromagnets by the multi-sublattice construction of
Section~\ref{sec:afmsw}. Thirteen of the fourteen lattices therefore carry a complete
spin-wave analysis, and only the incommensurate face-centred orthorhombic case lies
outside standard linear spin-wave theory.

\section{Validation}

Three published values, obtained by methods sharing no machinery with one another or with
ours, are reproduced to the precision at which they were quoted:
\begin{itemize}
\item \textbf{simple cubic}, $e=-2.676789$ at $\bk_0=(\pi,0,\pi)$, against $-2.67679$ of
Batle \& Ciftja~\cite{BatleCiftja2020}, obtained by finite-cluster energy decomposition with
neither Ewald summation nor Luttinger--Tisza analysis;
\item \textbf{face-centred cubic}, $e=-2.961922$, against the exact tinfoil value
$-2\pi\sqrt2/3=-2.9619219\ldots$ and the $-2.961921952$ of Gross \& Wei~\cite{GrossWei2000};
\item \textbf{body-centred tetragonal} at $c/a=\sqrt{2/3}$, $e=-3.050099878$, against
$-3.050099872$ of Ref.~\cite{GrossWei2000}, agreement to nine significant figures.
\end{itemize}
Internally, $\lambda_{\min}[\Am(\bk)]$ is independent of the Ewald parameter $\alpha$ to
$\sim5\times10^{-15}$ at generic $\bk$, and the rhombohedral cell at $60^{\circ}$
reproduces the fcc value exactly, as it must, being the same lattice in a different basis.

\section{Results}

Table~\ref{tab:main} collects the fourteen lattices, ordered by binding energy. Six order
ferromagnetically and eight at finite wavevector, and the split follows the
centring rule of Luttinger and Tisza~\cite{LuttingerTisza1946}: primitive lattices order at finite
$\bk$, body- and face-centred ones ferromagnetically.

\begin{table*}[t]
\centering
\small
\begin{tabular}{lrlrrr}
\toprule
lattice & $e_{\min}$ & $\mathbf k_0$ & $\delta S$ & $e_{\mathrm{qu}}$ & $V$ \\
\midrule
body-centred tetragonal tI ($c/a=\sqrt{2/3}$) & $-3.050100$ & $\Gamma$ & $0.1493$ & $-3.677319$ & $0.750$\\
face-centred cubic cF & $-2.961922$ & $\Gamma$ & $0.2446$ & $-3.764702$ & $0.707$\\
rhombohedral hR ($60^\circ$; $=$ fcc) & $-2.961922$ & $\Gamma$ & $0.2466$ & $-3.769300$ & $0.707$\\
body-centred orthorhombic oI & $-2.927209$ & $\Gamma$ & $0.1792$ & $-3.590392$ & $0.755$\\
hexagonal hP ($c/a=1.3$) & $-2.771222$ & $(0,0,\tfrac12)$ & $0.0593$ & $-3.027026$ & $1.126$\\
body-centred cubic cI & $-2.720699$ & $\Gamma$ & $0.2591$ & $-3.489567$ & $0.770$\\
simple cubic cP & $-2.676789$ & $(\tfrac12,0,\tfrac12)$ & $0.2649$ & $-3.053360$ & $1.000$\\
face-centred orthorhombic oF$^{\dagger}$ & $-2.627976$ & (0.15,0.15,0.30) & $0.1512$ & $-3.172943$ & $0.846$\\
base-centred orthorhombic oC & $-2.582432$ & $(0,0,\tfrac12)$ & $0.0279$ & $-2.714187$ & $1.570$\\
tetragonal tP ($c/a=1.3$) & $-2.564757$ & $(0,\tfrac12,\tfrac12)$ & $0.0789$ & $-2.696594$ & $1.300$\\
base-centred monoclinic mC & $-2.508821$ & $\Gamma$ & $0.0402$ & $-2.683099$ & $1.534$\\
triclinic aP & $-2.481090$ & $(0,0,\tfrac12)$ & $0.0307$ & $-2.622507$ & $1.208$\\
monoclinic mP & $-2.467658$ & $(0,\tfrac12,0)$ & $0.0180$ & $-2.547048$ & $1.438$\\
orthorhombic oP & $-2.448194$ & $(0,\tfrac12,\tfrac12)$ & $0.0102$ & $-2.493762$ & $1.740$\\
\bottomrule
\end{tabular}
\caption{Dipolar ground states of the fourteen Bravais lattices, nearest-neighbour distance
normalised to unity, tinfoil convention, one dipole per primitive cell. Free cell parameters
are fixed at the representative values listed in the panels. Spin-wave quantities are given
for \emph{all} lattices: by Eq.~\eqref{eq:magnon} for the ferromagnets, by the
multi-sublattice treatment of Section~\ref{sec:afmsw} for the commensurate collinear states,
and, for oF, about its true ground state (Section~\ref{sec:oF}). $e_{\mathrm{qu}}$ is the
magnon-corrected energy of Eq.~\eqref{eq:zeropoint}, so $e_{\mathrm{qu}}-e_{\min}$ is the
zero-point energy shift; it is a different quantity from $\delta S$, the zero-point
reduction of the ordered moment, and the two are not related by a simple sum.
$^{\dagger}$For face-centred orthorhombic alone the Luttinger--Tisza value $-2.627976$ is a
strict lower bound and is \emph{not} attained; its true ground-state energy is $-2.609170$,
and the $\delta S$ and $e_{\mathrm{qu}}$ in its row refer to that true state, not to the
Luttinger--Tisza state. $V$ is the primitive-cell volume.}
\label{tab:main}
\end{table*}

\begin{figure*}[tbp]
\centering
\includegraphics[width=\textwidth]{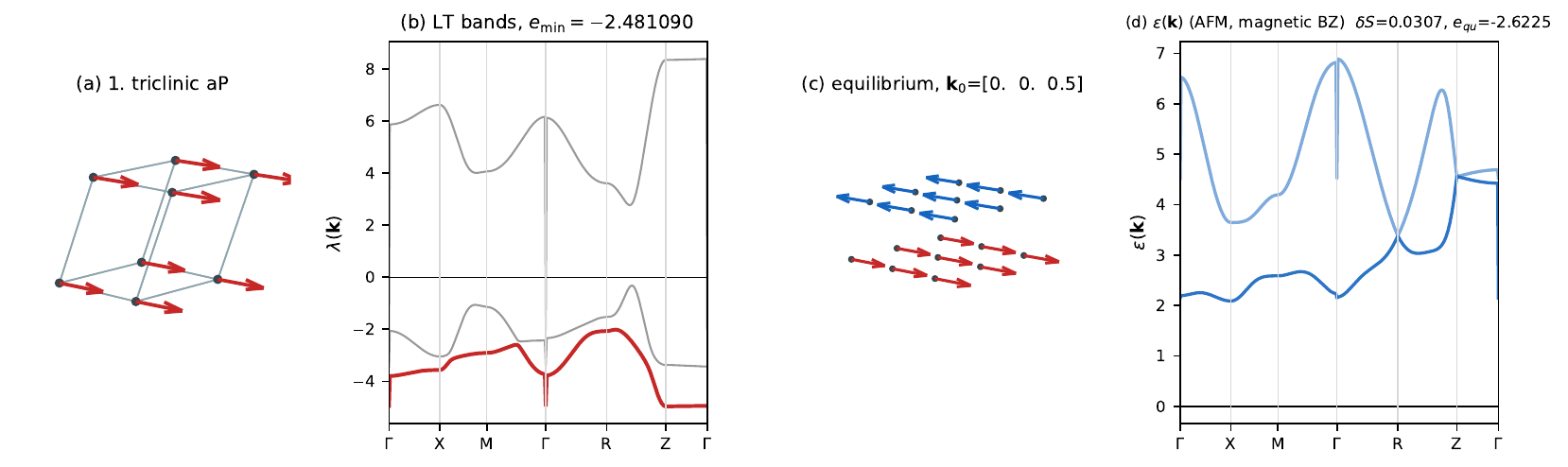}
\caption{triclinic aP. $e_{\min}=-2.481090$, finite $\mathbf k_0=(0.00,0.00,0.50)$. Panels as described in the text: (a) primitive cell with
equilibrium moments; (b) Luttinger--Tisza bands, lowest highlighted; (c) equilibrium
configuration; (d) magnon dispersion, computed over the magnetic Brillouin zone of the doubled cell, the ground state being a collinear antiferromagnet. Equilibrium orientations: $\phi=5^\circ$ and $185^\circ$, i.e.\ moments $\pm\hat n$ alternating along the axis selected by $\mathbf k_0$. In panel (c) the moments are drawn as $\cos(\mathbf k_0\cdot\mathbf R)\,\hat n$ and coloured by the sign of that factor: red where it is $+1$ and blue where it is $-1$. Red and blue are therefore the two antiparallel sublattices of a single antiferromagnet, not two separate structures; they appear here as alternating layers because $\mathbf k_0$ selects that stacking direction.}
\label{fig:p1}
\end{figure*}

\begin{figure*}[tbp]
\centering
\includegraphics[width=\textwidth]{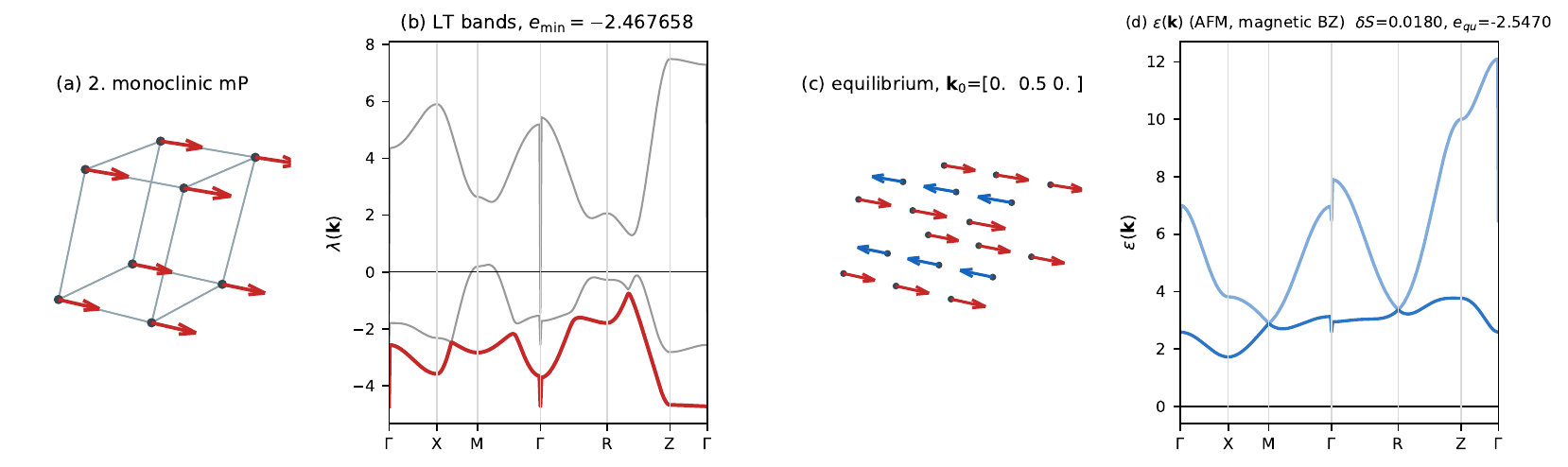}
\caption{monoclinic mP. $e_{\min}=-2.467658$, finite $\mathbf k_0=(0.00,0.50,0.00)$. Panels as described in the text: (a) primitive cell with
equilibrium moments; (b) Luttinger--Tisza bands, lowest highlighted; (c) equilibrium
configuration; (d) magnon dispersion, computed over the magnetic Brillouin zone of the doubled cell, the ground state being a collinear antiferromagnet. Equilibrium orientations: $\phi=0^\circ$ and $180^\circ$, i.e.\ moments $\pm\hat n$ alternating along the axis selected by $\mathbf k_0$. In panel (c) the moments are drawn as $\cos(\mathbf k_0\cdot\mathbf R)\,\hat n$ and coloured by the sign of that factor: red where it is $+1$ and blue where it is $-1$. Red and blue are therefore the two antiparallel sublattices of a single antiferromagnet, not two separate structures; they appear here as alternating layers because $\mathbf k_0$ selects that stacking direction.}
\label{fig:p2}
\end{figure*}

The ordering is \emph{not} by density: fcc, the densest packing
at $V=0.707$, is beaten by body-centred tetragonal at $V=0.750$. The dipolar problem
selects a chain-forming geometry rather than the closest packing, which is the essential
qualitative difference from the classical scalar lattice-energy problems. Second, the
finite-$\bk$ ground states are almost all at zone-boundary points with fractional
coordinates in $\{0,\tfrac12\}$, i.e.\ commensurate antiferromagnetic or stripe order; the
one exception is face-centred orthorhombic at $(0.15,0.15,0.30)$, an incommensurate state.
Finally, the two lowest entries at $-2.961922$ are the same lattice, which serves as an
internal consistency check. Finally, the magnon dispersions of panel (d) are discontinuous
where the path crosses $\Gamma$; this is a physical consequence of the non-analyticity of
$\Am(\bk)$ at long wavelength and is explained in Remark~\ref{rem:gammajump}.

\begin{figure*}[tbp]
\centering
\includegraphics[width=\textwidth]{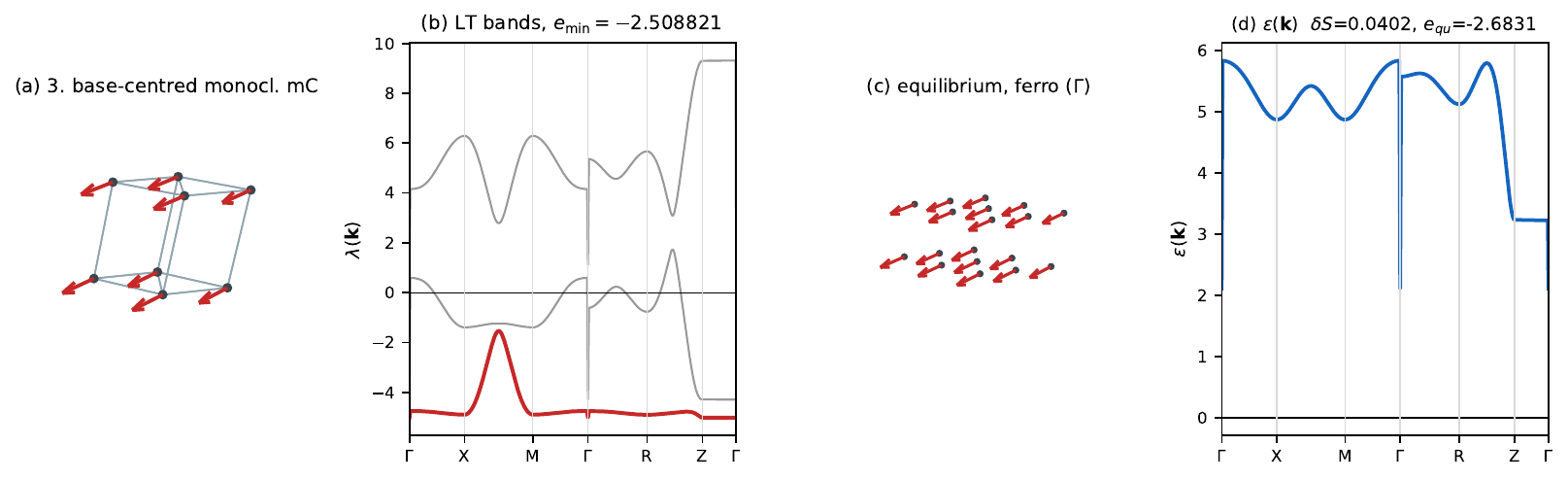}
\caption{base-centred monocl. mC. $e_{\min}=-2.508821$, ferromagnetic ($\mathbf k_0=\Gamma$). Panels as described in the text: (a) primitive cell with
equilibrium moments; (b) Luttinger--Tisza bands, lowest highlighted; (c) equilibrium
configuration; (d) magnon dispersion where the ground state is ferromagnetic. Equilibrium orientation: all moments parallel, easy axis at $(\theta,\phi)=(90^\circ,270^\circ)$ in the cell frame.}
\label{fig:p3}
\end{figure*}

\begin{figure*}[tbp]
\centering
\includegraphics[width=\textwidth]{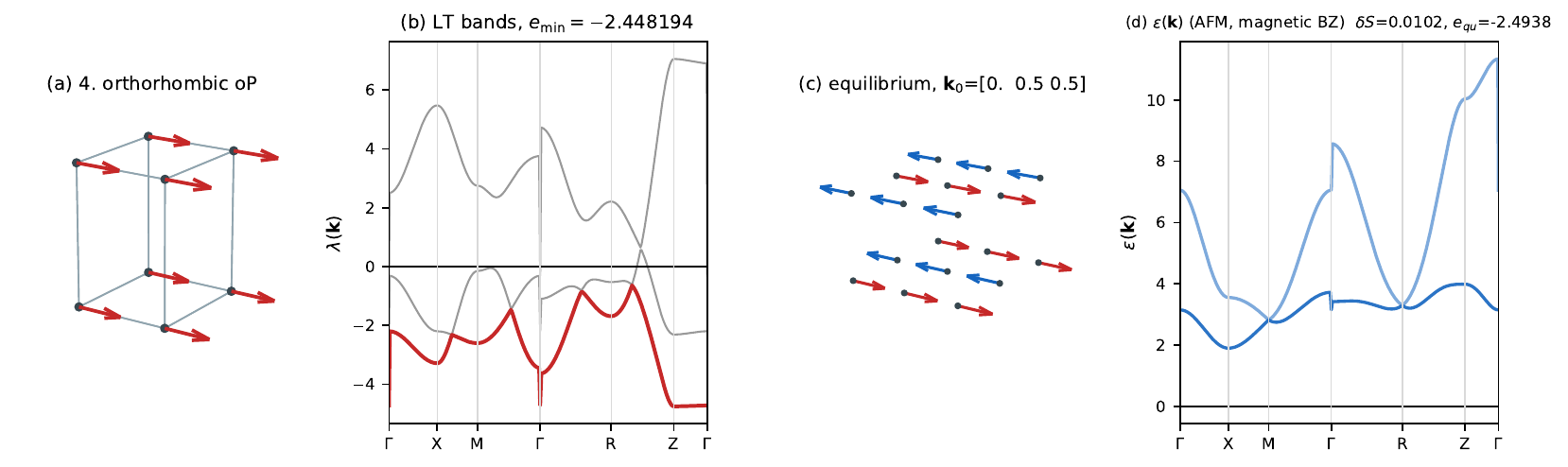}
\caption{orthorhombic oP. $e_{\min}=-2.448194$, finite $\mathbf k_0=(0.00,0.50,0.50)$. Panels as described in the text: (a) primitive cell with
equilibrium moments; (b) Luttinger--Tisza bands, lowest highlighted; (c) equilibrium
configuration; (d) magnon dispersion, computed over the magnetic Brillouin zone of the doubled cell, the ground state being a collinear antiferromagnet. Equilibrium orientations: $\phi=0^\circ$ and $180^\circ$, i.e.\ moments $\pm\hat n$ alternating along the axis selected by $\mathbf k_0$. In panel (c) the moments are drawn as $\cos(\mathbf k_0\cdot\mathbf R)\,\hat n$ and coloured by the sign of that factor: red where it is $+1$ and blue where it is $-1$. Red and blue are therefore the two antiparallel sublattices of a single antiferromagnet, not two separate structures; they appear here as alternating layers because $\mathbf k_0$ selects that stacking direction.}
\label{fig:p4}
\end{figure*}

\begin{figure*}[tbp]
\centering
\includegraphics[width=\textwidth]{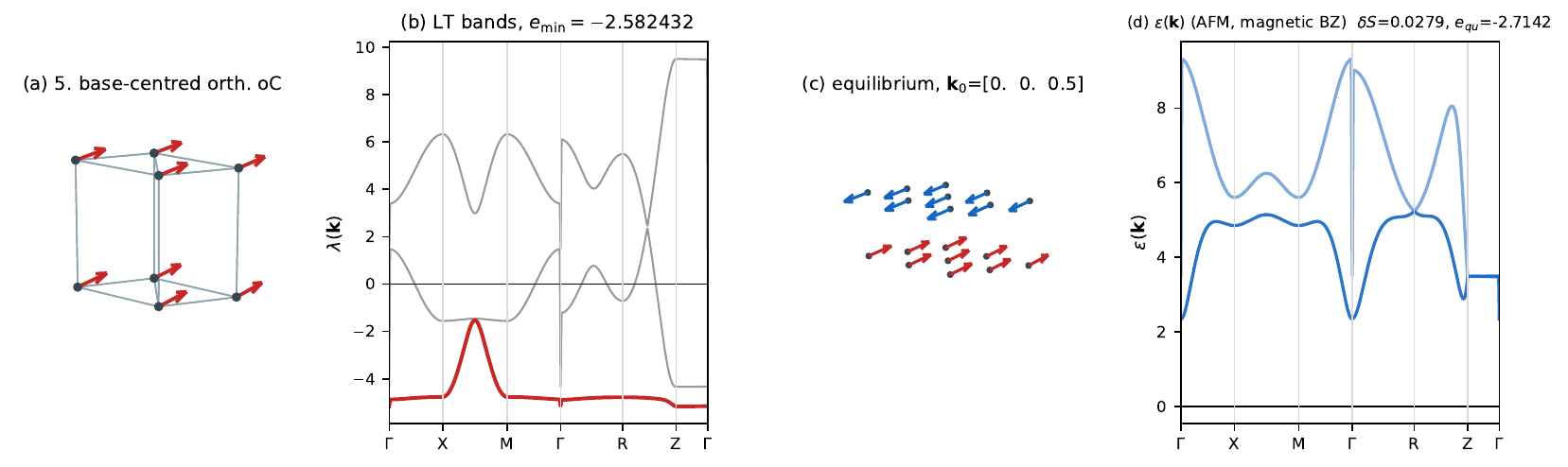}
\caption{base-centred orth. oC. $e_{\min}=-2.582432$, finite $\mathbf k_0=(0.00,0.00,0.50)$. Panels as described in the text: (a) primitive cell with
equilibrium moments; (b) Luttinger--Tisza bands, lowest highlighted; (c) equilibrium
configuration; (d) magnon dispersion, computed over the magnetic Brillouin zone of the doubled cell, the ground state being a collinear antiferromagnet. Equilibrium orientations: $\phi=90^\circ$ and $270^\circ$, i.e.\ moments $\pm\hat n$ alternating along the axis selected by $\mathbf k_0$. In panel (c) the moments are drawn as $\cos(\mathbf k_0\cdot\mathbf R)\,\hat n$ and coloured by the sign of that factor: red where it is $+1$ and blue where it is $-1$. Red and blue are therefore the two antiparallel sublattices of a single antiferromagnet, not two separate structures; they appear here as alternating layers because $\mathbf k_0$ selects that stacking direction.}
\label{fig:p5}
\end{figure*}

\subsection{Spin waves of the collinear antiferromagnets}\label{sec:afmsw}

Seven of the eight finite-$\bk$ ground states are commensurate, hence collinear, and are
therefore accessible to conventional multi-sublattice linear spin-wave theory. We compute
them here, so that thirteen of the fourteen lattices now carry a spin-wave analysis and
only the incommensurate face-centred orthorhombic case lies outside standard LSWT.

For $\bk_0$ with a half-integer fractional component, $\sigma_\bR=\cos(\bk_0\cdot\bR)=\pm1$
and the magnetic cell is the index-two sublattice
$\Lambda'=\{\bR:\bk_0\cdot\bR\in2\pi\mathbb Z\}$, carrying two cosets with moments
$+\hat n$ and $-\hat n$. With local frames $(\hat x_i,\hat y_i,\hat z_i)$, $u_i=\hat
x_i+i\hat y_i$, $v_i=\hat z_i$, the standard construction gives
\begin{equation}\begin{split}
A_{ij}(\bk)=\tfrac{S}{2}u_i\Am_{ij}(\bk)\bar u_j,\quad
B_{ij}(\bk)=\tfrac{S}{2}u_i\Am_{ij}(\bk)u_j,
\\[2pt]
C_{ij}=\delta_{ij}S\sum_l v_i\Am_{il}(0)v_l,
\end{split}\end{equation}
and the magnon energies are the positive eigenvalues of $gH(\bk)$ with
\begin{equation}
H(\bk)=\begin{pmatrix}A-C & B\\ B^{\dagger} & \bar A(-\bk)-C\end{pmatrix},
\quad g=\mathrm{diag}(I,-I),
\end{equation}
diagonalised by Colpa's method~\cite{Colpa1978}; the general
multi-sublattice formulation we follow is that of Toth and Lake~\cite{TothLake2015}. Applied to a single sublattice
with $\hat z=\hat n$ this reproduces the ferromagnetic branch of Eq.~\eqref{eq:magnon} to
$10^{-12}$, which validates the implementation.

\begin{table*}[t]
\centering\small
\begin{tabular}{lrrr}
\toprule
lattice (collinear AFM/stripe) & $e_{\min}$ & $\delta S$ & $e_{\mathrm{qu}}$\\
\midrule
simple cubic cP & $-2.676789$ & $0.2649$ & $-3.053360$ \\
hexagonal hP c/a=1.3 & $-2.771222$ & $0.0593$ & $-3.027026$ \\
tetragonal tP c/a=1.3 & $-2.564757$ & $0.0789$ & $-2.696594$ \\
orthorhombic oP & $-2.448194$ & $0.0102$ & $-2.493762$ \\
base-centred orth. oC & $-2.582432$ & $0.0279$ & $-2.714187$ \\
monoclinic mP & $-2.467658$ & $0.0180$ & $-2.547048$ \\
triclinic aP & $-2.481090$ & $0.0307$ & $-2.622507$ \\
\bottomrule
\end{tabular}
\caption{Spin-wave results for the commensurate collinear ground states. Energies are
per site. Wavevectors at which the Bogoliubov factorisation is singular --- Goldstone
points, a set of measure zero --- are excluded from the zone averages. The zero-point
reductions are of the same order as the planar value $0.0563$ for the square lattice.}
\label{tab:afmsw}
\end{table*}

Two remarks. First, simple cubic carries a substantially larger zero-point reduction than
the other collinear states, $\delta S=0.265$ against $0.010$--$0.079$, and a correspondingly
large fraction of its zone shows singular Bogoliubov factorisation. This is the signature of
the well-known continuous degeneracy of the simple-cubic dipolar ground state, present
already in Luttinger and Tisza's analysis: soft directions in configuration space produce
low-lying modes and hence strong quantum fluctuations. Second, and more informatively, the audit of Table~\ref{tab:main} shows that $\delta S$
does not separate ferromagnets from antiferromagnets. It separates lattices with a
degenerate easy axis from those without. The four cubic-symmetric entries --- sc, bcc, fcc
and its rhombohedral copy --- have $\delta S=0.245$--$0.265$, and by
Proposition~\ref{prop:cubic} all have $\Am(0)\equiv0$, hence no anisotropy and a completely
degenerate easy direction. The three weakly anisotropic entries (bct, oI, oF) follow at
$0.149$--$0.179$, and the genuinely anisotropic low-symmetry lattices come last at
$0.010$--$0.079$, the range familiar from the planar survey. Soft directions in
configuration space produce low-lying modes and hence strong quantum fluctuations; the
ordering of the $\delta S$ column is a direct measure of how much anisotropy each lattice
possesses.

\begin{figure*}[tbp]
\centering
\includegraphics[width=\textwidth]{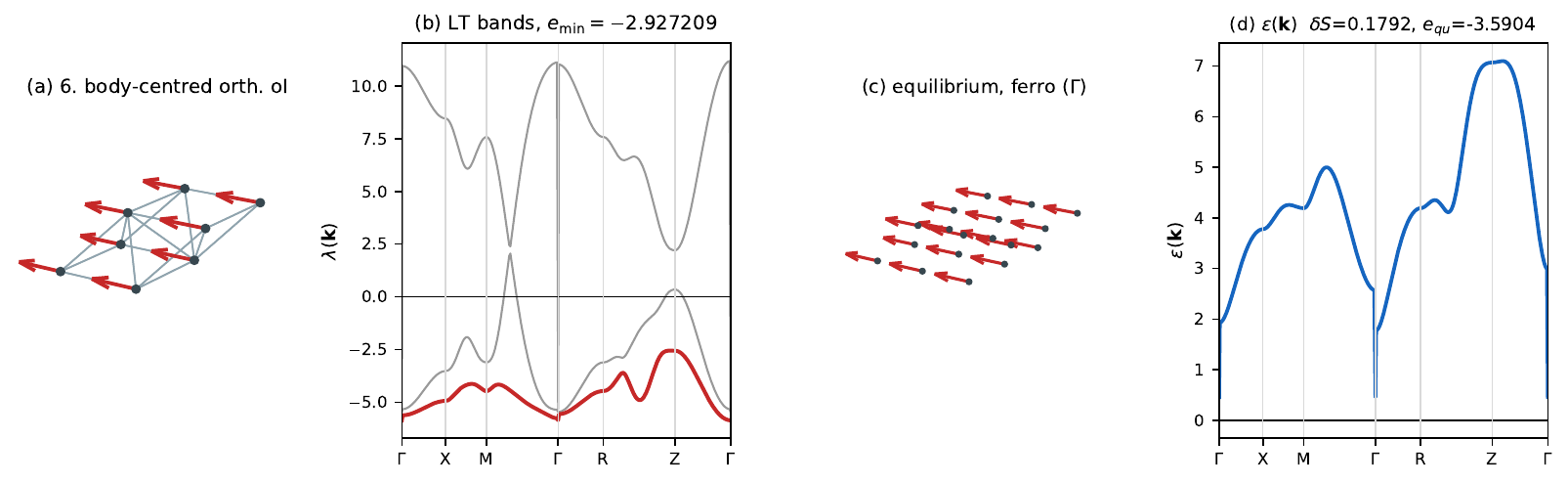}
\caption{body-centred orth. oI. $e_{\min}=-2.927209$, ferromagnetic ($\mathbf k_0=\Gamma$). Panels as described in the text: (a) primitive cell with
equilibrium moments; (b) Luttinger--Tisza bands, lowest highlighted; (c) equilibrium
configuration; (d) magnon dispersion where the ground state is ferromagnetic. Equilibrium orientation: all moments parallel, easy axis at $(\theta,\phi)=(90^\circ,180^\circ)$ in the cell frame.}
\label{fig:p6}
\end{figure*}

\begin{figure*}[tbp]
\centering
\includegraphics[width=\textwidth]{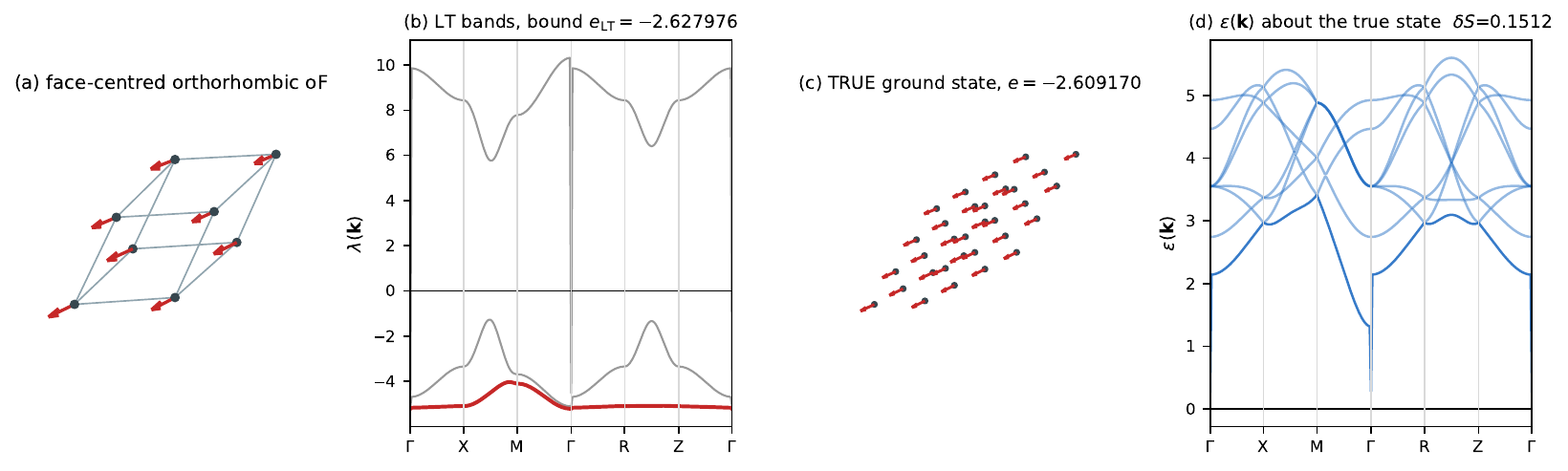}
\caption{Face-centred orthorhombic oF, the one lattice whose Luttinger--Tisza state is not
realisable. (a) primitive cell; (b) LT bands --- exact, and their minimum
$e_{\rm LT}=-2.627976$ is a rigorous lower bound, though not an attained energy; (c) the
\emph{true} ground state from direct supercell minimisation, $e=-2.609170$, all moments of
unit length; (d) its magnon spectrum about that state, eight branches, real and positive
throughout, $\delta S=0.1512$. See Section~\ref{sec:oF}.}
\label{fig:p7}
\end{figure*}

\subsection{The incommensurate lattice: why no spin-wave expansion exists}\label{sec:oF}

The argument runs in four steps, worth stating before the details. (i) The Luttinger--Tisza
construction is a relaxation, so its value is always a \emph{lower bound} on the true energy.
(ii) For this lattice the bound is not attained: the minimising eigenvector is not circular, so
the single-$\bk$ state is a spin-density wave whose moment length varies from site to site and
which therefore violates the hard constraint $|\mathbf S_\bR|=1$. (iii) There is consequently
no spiral to expand about, and every commensurate approximant we construct is dynamically
unstable --- the instability is the diagnostic, not a numerical failure. (iv) Abandoning the
single-$\bk$ ansatz and minimising directly over unit moments in a supercell yields a
converged, stable, non-collinear ground state lying above the bound.

Face-centred orthorhombic is the one lattice whose ordering wavevector,
$\bk_0=(0.1527,0.1491,0.3018)$, has no half-integer component. We attempted the
rotating-frame treatment appropriate to a helix, using commensurate approximants: replacing
$\bk_0$ by a rational $\mathbf Q=(a,b,c)/q$ makes
$\Lambda'=\{\bR:\,a n_1+b n_2+c n_3\equiv0\ (\mathrm{mod}\ q)\}$ a sublattice of index $q$,
whose $q$ cosets carry moments at angles $2\pi m/q$ in the spiral plane, and the
multi-sublattice machinery of Section~\ref{sec:afmsw} then applies with $q$ sublattices.
Every approximant tried, $q=3,5,6,7,10$, came out dynamically unstable.

The reason is not numerical, and it is more interesting than a dispersion would have been.
A single-$\bk$ state $\mathbf S_\bR=\mathrm{Re}(\mathbf m\,e^{i\bk_0\cdot\bR})$ has constant
length --- and is therefore an admissible configuration of unit moments --- only if
$\mathbf m$ is \emph{circular}, $\mathbf m=(\hat u+i\hat v)/\sqrt2$ with $\hat u\perp\hat v$
of equal magnitude. That requires $\lambda_{\min}[\Am(\bk_0)]$ to be degenerate. Here it is
not: the eigenvalues are $(-5.255952,\,-3.646662,\,8.902614)$, a gap of $1.609$. The
minimising eigenvector is therefore real up to an overall phase, and the Luttinger--Tisza
state is
\begin{equation}
\mathbf S_\bR=\hat n\,\cos(\bk_0\cdot\bR),
\end{equation}
a sinusoidal spin-density wave whose length varies from site to site and which violates
$|\mathbf S_\bR|=1$.

Consequently the entry $-2.627976$ in Table~\ref{tab:main} is, for this lattice alone, a
strict \emph{lower bound} rather than an attained energy: the Luttinger--Tisza optimum is
not realisable, the true ground state is some multi-$\bk$ or modulated structure lying
above it, and there is no spiral to expand about --- which is precisely why every
approximant is unstable. The instability is the diagnostic, not a failure of the method.

\subsubsection*{The true ground state, by direct supercell minimisation}

Determining the actual face-centred orthorhombic ground state requires abandoning the
single-$\bk$ ansatz altogether. We impose periodic boundary conditions on a supercell of
$N=n_1n_2n_3$ primitive cells, Ewald-sum the dipolar tensor for that supercell to obtain the
exact $N\times N$ block coupling, and minimise
\begin{equation}
\frac{E}{N}=\frac{1}{2N}\sum_{ij}\mathbf S_i\cdot\bm T_{ij}\cdot\mathbf S_j,
\qquad |\mathbf S_i|=1\ \ \forall i,
\end{equation}
over the $2N$ spherical angles by quasi-Newton descent from many annealed random starts. Basis reduction, used to fix the
nearest-neighbour normalisation for skewed cells, follows Lenstra, Lenstra and
Lov\'asz~\cite{LLL1982}.
Any configuration commensurate with the supercell is reachable, so several supercell shapes
test whether the answer is converged rather than an artefact of one cell. The procedure is
validated on body-centred tetragonal, where it returns $-3.050099878$ exactly, both for the
uniform state and after relaxation from a perturbed start.

\begin{table}[t]
\centering\small
\begin{tabular}{lrrl}
\toprule
supercell & $N$ & $e$ & character\\
\midrule
$2\times2\times2$ & $8$  & $-2.609170$ & non-collinear\\
$3\times3\times2$ & $18$ & $-2.609170$ & non-collinear\\
$3\times3\times3$ & $27$ & $-2.609170$ & non-collinear\\
$4\times4\times2$ & $32$ & $-2.609170$ & non-collinear\\
\bottomrule
\end{tabular}
\caption{Direct supercell minimisation for face-centred orthorhombic. Four supercell shapes
and sizes give the same energy to six decimals, indicating a converged result rather than a
commensuration artefact.}
\label{tab:ofgs}
\end{table}

The result is
\begin{equation}
e_{\mathrm{oF}}=-2.609170,\qquad
e_{\mathrm{oF}}-e_{\mathrm{LT}}=+0.018806,
\end{equation}
so the hard-spin constraint costs $0.0188$, about $0.7\%$, relative to the unattainable
Luttinger--Tisza optimum. The ground state is non-collinear, as the failure of the
single-$\bk$ construction requires. Face-centred orthorhombic is thus the only lattice of
the fourteen whose tabulated Luttinger--Tisza energy is a bound rather than an attained
value, and the gap is now quantified rather than merely flagged.

\subsubsection*{What remains valid, and what Figure~\ref{fig:p7} shows}

It is worth being explicit about what the failure of the single-$\bk$ construction does and
does not invalidate, since the point is easily misread.

The Luttinger--Tisza \emph{bands} are unaffected. They are the eigenvalues of the exactly
Ewald-summed $\Am(\bk)$, a spectral property of the lattice defined independently of any
configuration, and panel (b) of Figure~\ref{fig:p7} is as valid for face-centred
orthorhombic as for any other lattice. What fails is only the inference that the minimum of
the lowest band is the ground-state energy, which requires the minimising eigenvector to
satisfy $|\mathbf S_\bR|=1$. By the relaxation property of the construction, $e_{\rm LT}$
remains a rigorous lower bound over \emph{all} configurations, so the band minimum retains
a precise meaning; it is a bound, not an attained value.

Panels (c) and (d) are therefore drawn about the \emph{true} ground state rather than the
Luttinger--Tisza state: (c) shows the relaxed configuration of energy $-2.609170$, and (d)
its magnon spectrum, obtained by applying the multi-sublattice theory of
Section~\ref{sec:afmsw} with the eight sites of the minimising supercell as sublattices.
The spectrum has eight branches, is real and positive at every wavevector sampled along the
path, and yields $\delta S=0.1512$ and $e_{\mathrm{qu}}=-3.172943$, i.e.\ a zero-point energy shift
$e_{\mathrm{qu}}-e=-0.563773$. (The two numbers measure different things:
$\delta S$ is the reduction of the ordered moment, $e_{\mathrm{qu}}-e$ the shift in energy;
neither is obtained from the other by addition.) That the state is
dynamically stable confirms it as a genuine local minimum, in contrast to the
Luttinger--Tisza state, about which no stable expansion exists. No panel of the figure
displays a configuration that does not exist, and no analysis is lost.

\subsection{Equilibrium orientations}\label{sec:angles}

Following the convention of the planar survey we record, for each lattice, the equilibrium
orientation explicitly. For the ferromagnets every moment shares one direction, quoted as
the polar and azimuthal angles $(\theta,\phi)$ of the easy axis in the cell frame; for the
collinear states the two sublattices carry $\pm\hat n$, i.e.\ azimuths $\phi$ and
$\phi+180^\circ$, alternating along the axis singled out by $\bk_0$. These angles are given in the caption of each
figure. For the cubic lattices the angles are \emph{not}
meaningful: by Proposition~\ref{prop:cubic} the $\bk=0$ tensor vanishes identically, every
direction is degenerate, and the value reported is an arbitrary representative of that
degenerate manifold.

\begin{figure*}[tbp]
\centering
\includegraphics[width=\textwidth]{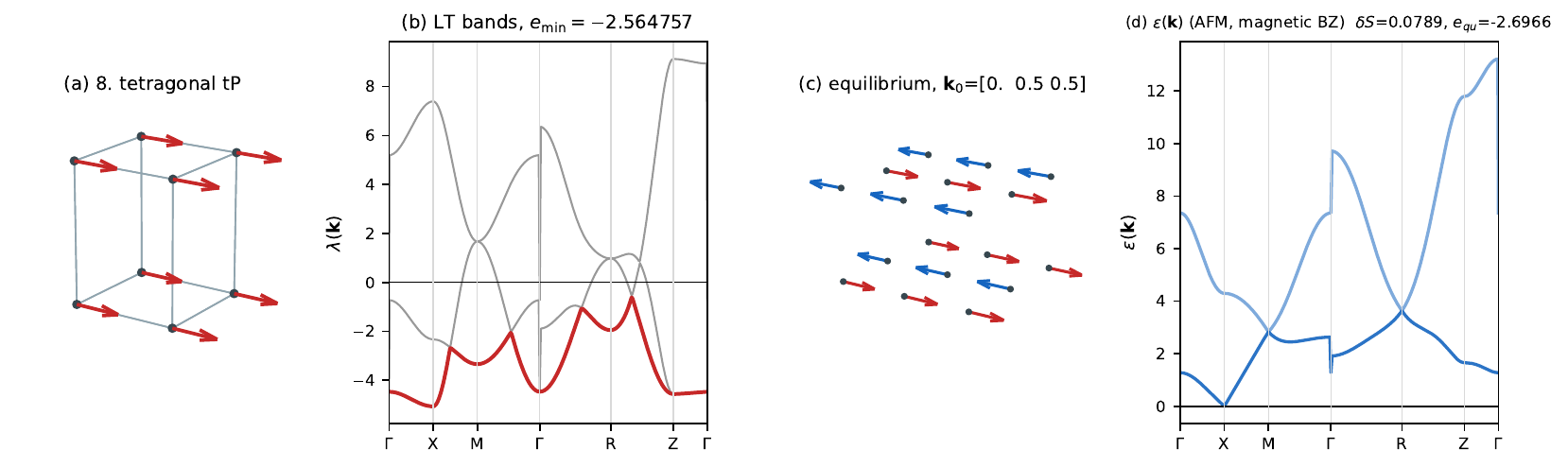}
\caption{tetragonal tP. $e_{\min}=-2.564757$, finite $\mathbf k_0=(0.00,0.50,0.50)$. Panels as described in the text: (a) primitive cell with
equilibrium moments; (b) Luttinger--Tisza bands, lowest highlighted; (c) equilibrium
configuration; (d) magnon dispersion, computed over the magnetic Brillouin zone of the doubled cell, the ground state being a collinear antiferromagnet. Equilibrium orientations: $\phi=0^\circ$ and $180^\circ$, i.e.\ moments $\pm\hat n$ alternating along the axis selected by $\mathbf k_0$. In panel (c) the moments are drawn as $\cos(\mathbf k_0\cdot\mathbf R)\,\hat n$ and coloured by the sign of that factor: red where it is $+1$ and blue where it is $-1$. Red and blue are therefore the two antiparallel sublattices of a single antiferromagnet, not two separate structures; they appear here as alternating layers because $\mathbf k_0$ selects that stacking direction.}
\label{fig:p8}
\end{figure*}

\begin{figure*}[tbp]
\centering
\includegraphics[width=\textwidth]{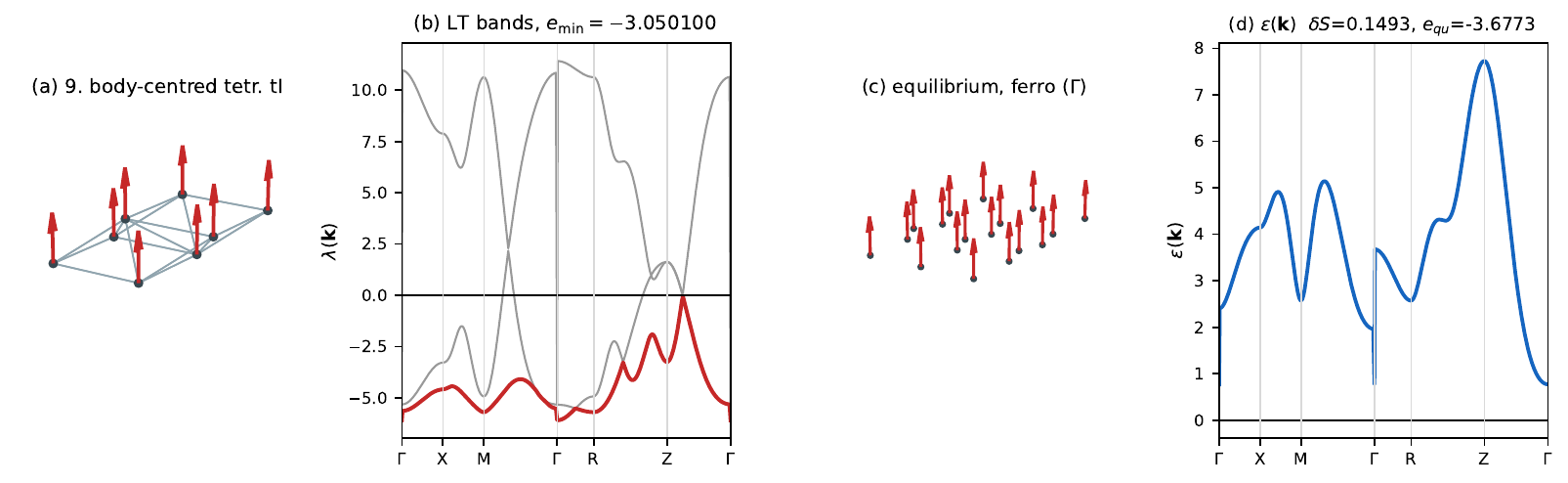}
\caption{body-centred tetr. tI. $e_{\min}=-3.050100$, ferromagnetic ($\mathbf k_0=\Gamma$). Panels as described in the text: (a) primitive cell with
equilibrium moments; (b) Luttinger--Tisza bands, lowest highlighted; (c) equilibrium
configuration; (d) magnon dispersion where the ground state is ferromagnetic. Equilibrium orientation: all moments parallel, easy axis at $(\theta,\phi)=(0^\circ,90^\circ)$ in the cell frame.}
\label{fig:p9}
\end{figure*}

\begin{figure*}[tbp]
\centering
\includegraphics[width=\textwidth]{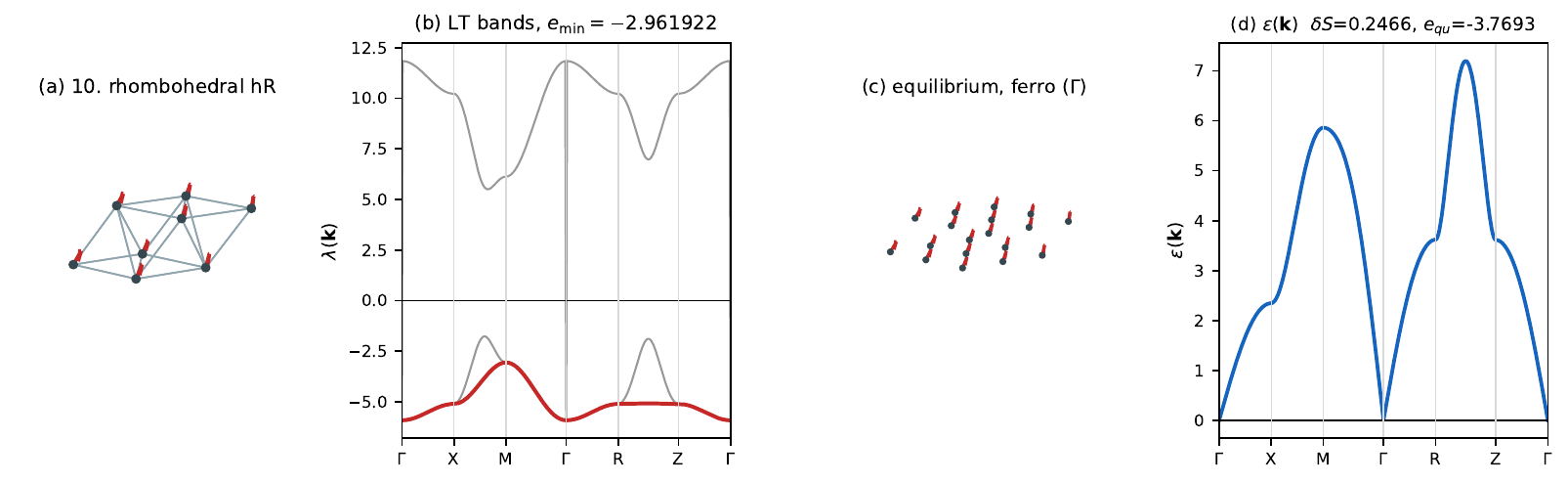}
\caption{rhombohedral hR. $e_{\min}=-2.961922$, ferromagnetic ($\mathbf k_0=\Gamma$). Panels as described in the text: (a) primitive cell with
equilibrium moments; (b) Luttinger--Tisza bands, lowest highlighted; (c) equilibrium
configuration; (d) magnon dispersion where the ground state is ferromagnetic. Equilibrium orientation: all moments parallel, easy axis at $(\theta,\phi)=(87^\circ,118^\circ)$ in the cell frame. At $\alpha=60^\circ$ the rhombohedral cell \emph{is} the face-centred cubic lattice in a different basis; this figure and Fig.~\ref{fig:p14} therefore describe the same structure, and their agreement to all quoted digits serves as an internal consistency check.}
\label{fig:p10}
\end{figure*}

\begin{figure*}[tbp]
\centering
\includegraphics[width=\textwidth]{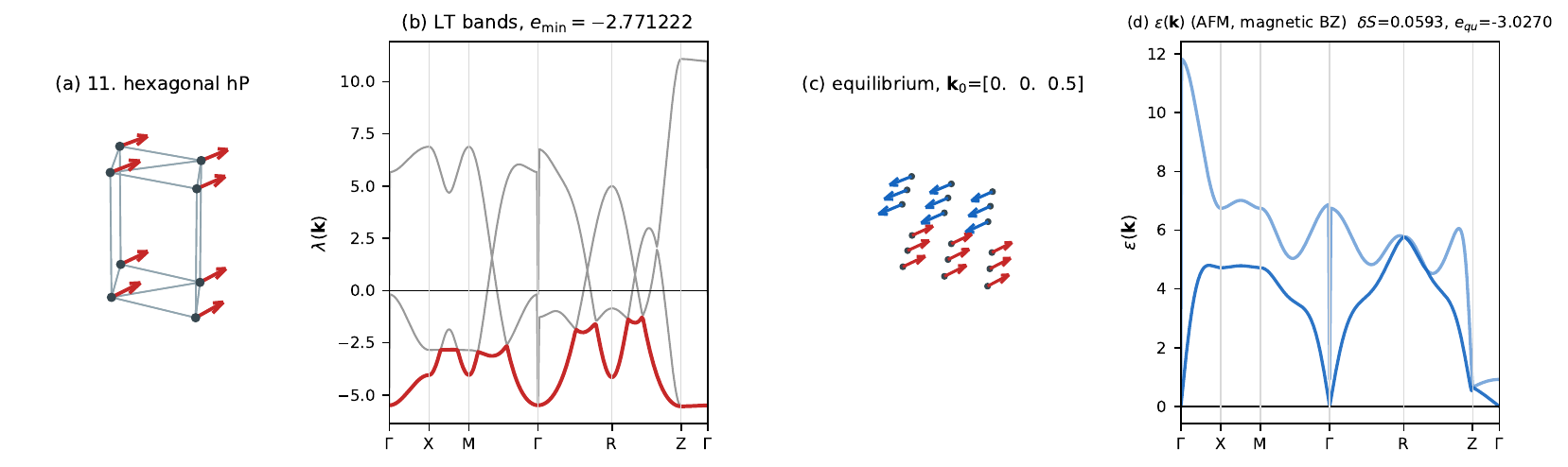}
\caption{hexagonal hP. $e_{\min}=-2.771222$, finite $\mathbf k_0=(0.00,0.00,0.50)$. Panels as described in the text: (a) primitive cell with
equilibrium moments; (b) Luttinger--Tisza bands, lowest highlighted; (c) equilibrium
configuration; (d) magnon dispersion, computed over the magnetic Brillouin zone of the doubled cell, the ground state being a collinear antiferromagnet. Equilibrium orientations: $\phi=90^\circ$ and $270^\circ$, i.e.\ moments $\pm\hat n$ alternating along the axis selected by $\mathbf k_0$. In panel (c) the moments are drawn as $\cos(\mathbf k_0\cdot\mathbf R)\,\hat n$ and coloured by the sign of that factor: red where it is $+1$ and blue where it is $-1$. Red and blue are therefore the two antiparallel sublattices of a single antiferromagnet, not two separate structures; they appear here as alternating layers because $\mathbf k_0$ selects that stacking direction.}
\label{fig:p11}
\end{figure*}

\begin{figure*}[tbp]
\centering
\includegraphics[width=\textwidth]{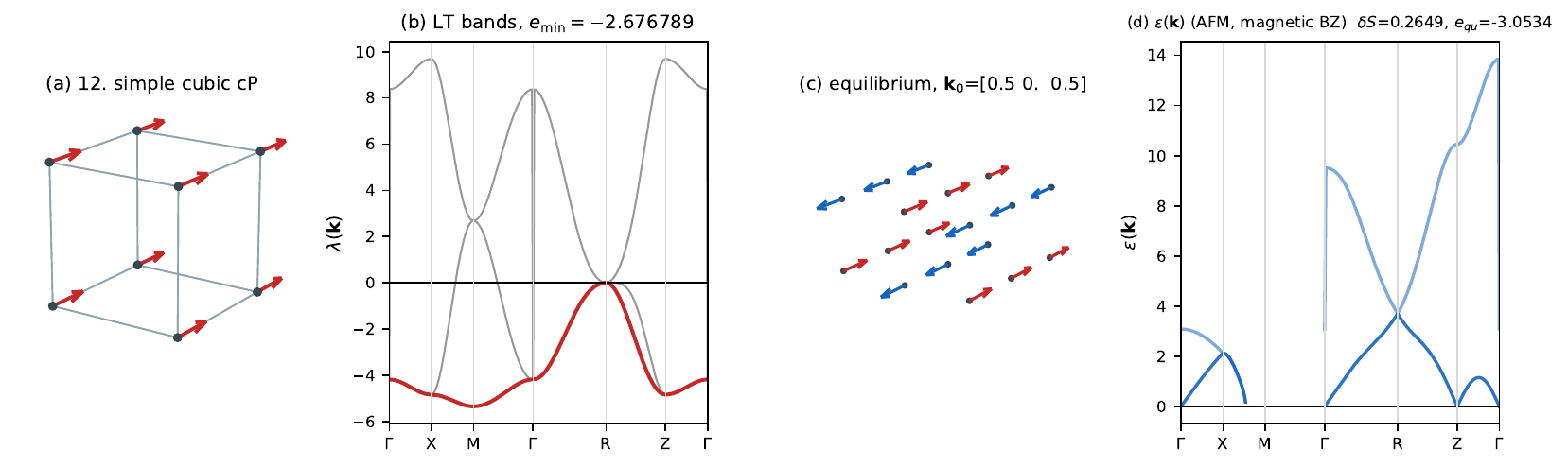}
\caption{simple cubic cP. $e_{\min}=-2.676789$, finite $\mathbf k_0=(0.50,0.00,0.50)$. Panels as described in the text: (a) primitive cell with
equilibrium moments; (b) Luttinger--Tisza bands, lowest highlighted; (c) equilibrium
configuration; (d) magnon dispersion, computed over the magnetic Brillouin zone of the doubled cell, the ground state being a collinear antiferromagnet. Equilibrium orientations: $\phi=90^\circ$ and $270^\circ$, i.e.\ moments $\pm\hat n$ alternating along the axis selected by $\mathbf k_0$. In panel (c) the moments are drawn as $\cos(\mathbf k_0\cdot\mathbf R)\,\hat n$ and coloured by the sign of that factor: red where it is $+1$ and blue where it is $-1$. Red and blue are therefore the two antiparallel sublattices of a single antiferromagnet, not two separate structures; they appear here as alternating layers because $\mathbf k_0$ selects that stacking direction.}
\label{fig:p12}
\end{figure*}

\begin{figure*}[tbp]
\centering
\includegraphics[width=\textwidth]{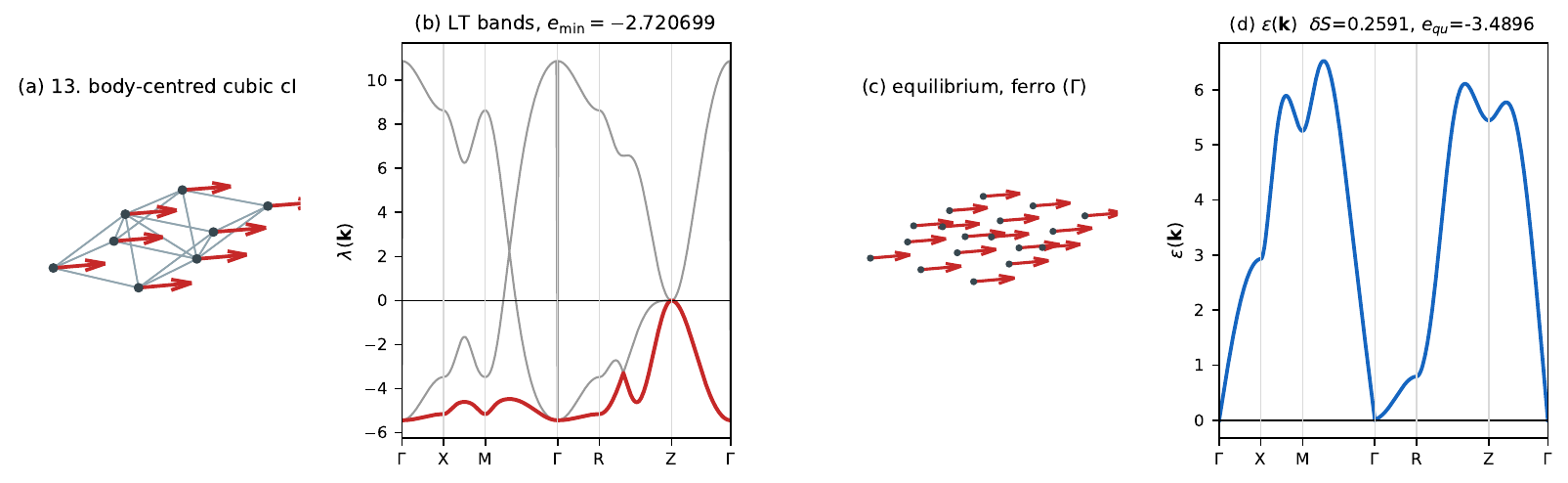}
\caption{body-centred cubic cI. $e_{\min}=-2.720699$, ferromagnetic ($\mathbf k_0=\Gamma$). Panels as described in the text: (a) primitive cell with
equilibrium moments; (b) Luttinger--Tisza bands, lowest highlighted; (c) equilibrium
configuration; (d) magnon dispersion where the ground state is ferromagnetic. Equilibrium orientation: all moments parallel, easy axis at $(\theta,\phi)=(90^\circ,48^\circ)$ in the cell frame.}
\label{fig:p13}
\end{figure*}

\begin{figure*}[tbp]
\centering
\includegraphics[width=\textwidth]{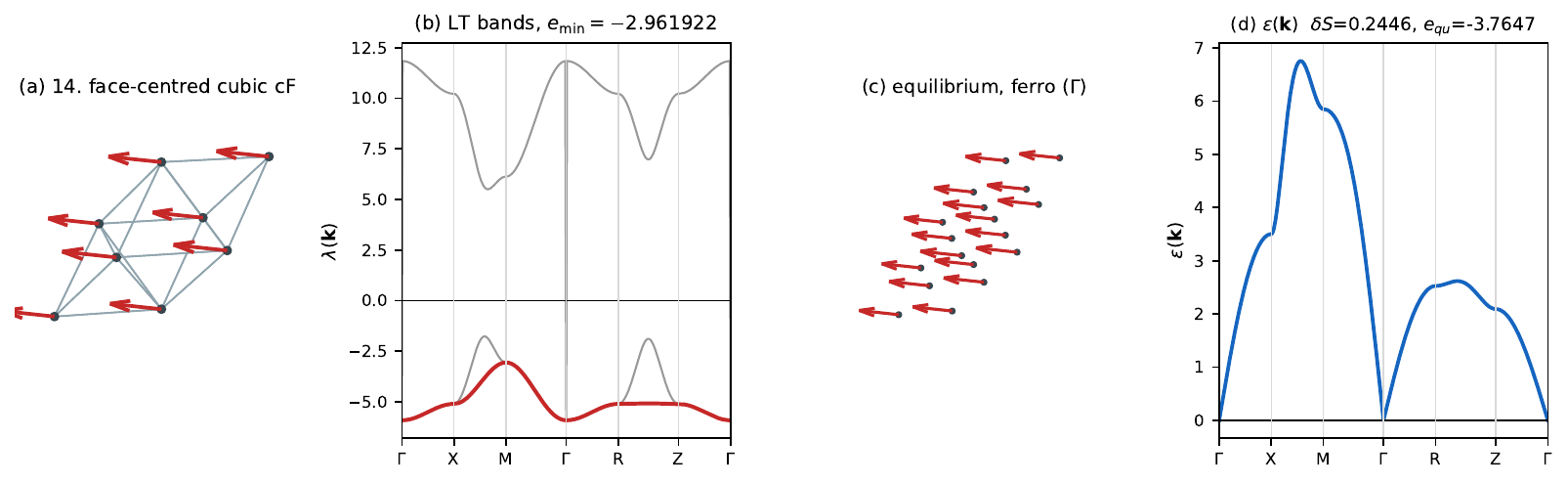}
\caption{face-centred cubic cF. $e_{\min}=-2.961922$, ferromagnetic ($\mathbf k_0=\Gamma$). Panels as described in the text: (a) primitive cell with
equilibrium moments; (b) Luttinger--Tisza bands, lowest highlighted; (c) equilibrium
configuration; (d) magnon dispersion where the ground state is ferromagnetic. Equilibrium orientation: all moments parallel, easy axis at $(\theta,\phi)=(85^\circ,210^\circ)$ in the cell frame.}
\label{fig:p14}
\end{figure*}

\section{Optimal member of each family: the phase diagram}\label{sec:optima}

Table~\ref{tab:main} reports representative geometries. We now optimise the free metric
parameters within each family, which converts the catalogue into a phase diagram.

One structural caveat governs the exercise. The fourteen families are \emph{not} disjoint:
each low-symmetry family contains higher-symmetry members in its closure, so an
unconstrained minimisation inside such a family migrates to the boundary and returns not a
distinct ``optimal triclinic'' but whichever higher-symmetry structure the closure reaches.
That is not a defect of the method; it is the correct answer to the question asked, and it
is what makes the resulting picture simple.

\begin{table*}[t]
\centering\small
\begin{tabular}{llrl}
\toprule
family (free parameters) & optimum & $e_{\min}$ & identification \\
\midrule
body-centred tetragonal (c/a) & $c/a=0.81650$ & $-3.050100$ & $\sqrt{2/3}$: the global optimum\\
body-centred orthorhombic (b/a,c/a) & $b/a=c/a=1.2247$ & $-3.050100$ & $b=c$: is bct\\
face-centred orthorhombic (b/a,c/a) & $b/a=c/a=1.7321$ & $-3.050100$ & fct $\equiv$ bct\\
rhombohedral ($\alpha$) & $\alpha=62.4219^\circ$ & $-2.978125$ & \textbf{interior optimum}\\
base-centred orthorhombic (b/a,c/a) & $b/a=\sqrt3,\;c/a=1$ & $-2.871146$ & is simple hexagonal\\
hexagonal (c/a) & $c/a=1.0000$ & $-2.871146$ & simple hexagonal\\
tetragonal (c/a) & $c/a=1.0000$ & $-2.676789$ & is simple cubic\\
orthorhombic (b/a,c/a) & $b/a=c/a=1$ & $-2.676789$ & is simple cubic\\
cubic cP, cI, cF & --- & fixed & no free parameters\\
\bottomrule
\end{tabular}
\caption{Optimal member of each Bravais family with free metric parameters. Every family
whose closure contains bct flows to it; the rest terminate on one of four attractors. The
recovery of $c/a=\sqrt{2/3}$ to five decimals is an independent check of the optimiser
against the known result.}
\label{tab:optima}
\end{table*}

Under the family-wise optimisation described above, the metric space of Bravais lattices
drains into four limiting structures,
\begin{equation}\begin{split}
\underbrace{-3.050100}_{\text{bct}}\;<\;
\underbrace{-2.978125}_{\text{rhombohedral }62.42^\circ}
\\[2pt]
<\;\underbrace{-2.871146}_{\text{simple hexagonal}}\;<\;
\underbrace{-2.676789}_{\text{simple cubic}} .
\end{split}\end{equation}
Three families (tI, oI, oF) reach bct; two (oC, hP) reach simple hexagonal; two (tP, oP)
reach simple cubic. Only the rhombohedral family possesses a \emph{genuinely interior}
optimum, at $\alpha=62.42^\circ$, which lies below the face-centred cubic value
$-2.961922$ attained at $\alpha=60^\circ$ and is, to our knowledge, not previously noted.
That the optimiser independently recovers $c/a=\sqrt{2/3}$ for bct validates it against
the one case where the answer is known.

\begin{figure*}[t]
\centering
\includegraphics[width=\textwidth]{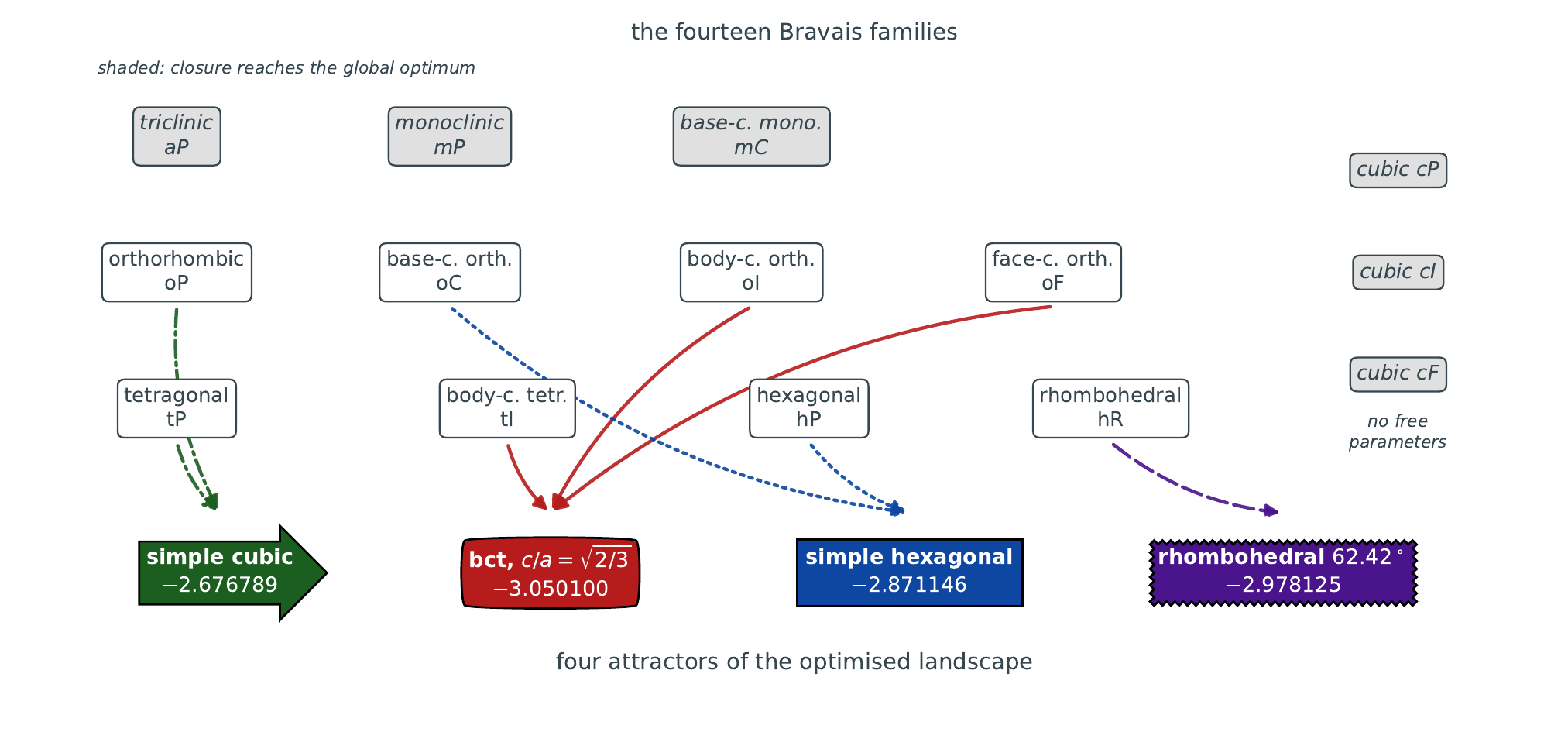}
\caption{Optimisation of the free metric parameters maps the fourteen Bravais families onto
four limiting structures. Box shape identifies the attractor (arrow, rounded, square and
sawtooth respectively), and arrow style matches it, so the grouping survives greyscale
reproduction. Shaded family labels denote lattices that are not optimisable in the sense used
here; the cubic lattices have no free parameters.}
\label{fig:attractors}
\end{figure*}

\begin{figure*}[p]
\centering
\includegraphics[width=\textwidth]{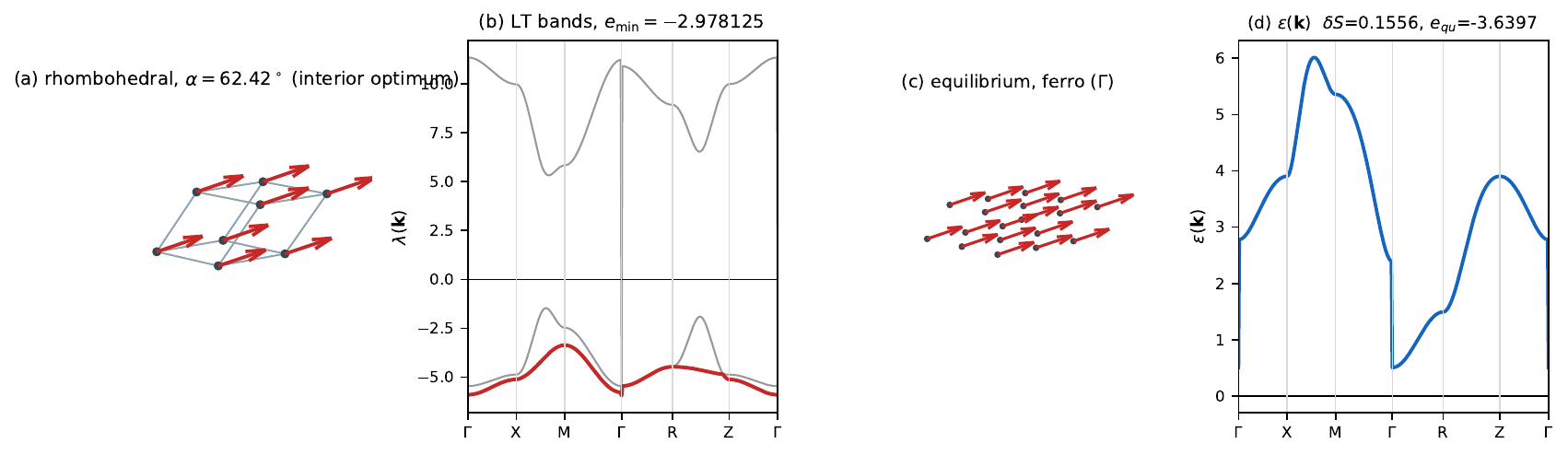}
\caption{The rhombohedral attractor: rhombohedral, $\alpha=62.42^\circ$ (interior optimum). $e_{\min}=-2.978125$, ferromagnetic ($\mathbf k_0=\Gamma$). $\delta S=0.1556$, $e_{\mathrm{qu}}=-3.639653$. Panels as
in Section~\ref{sec:panels}.}
\label{fig:at2}
\end{figure*}

\begin{figure*}[p]
\centering
\includegraphics[width=\textwidth]{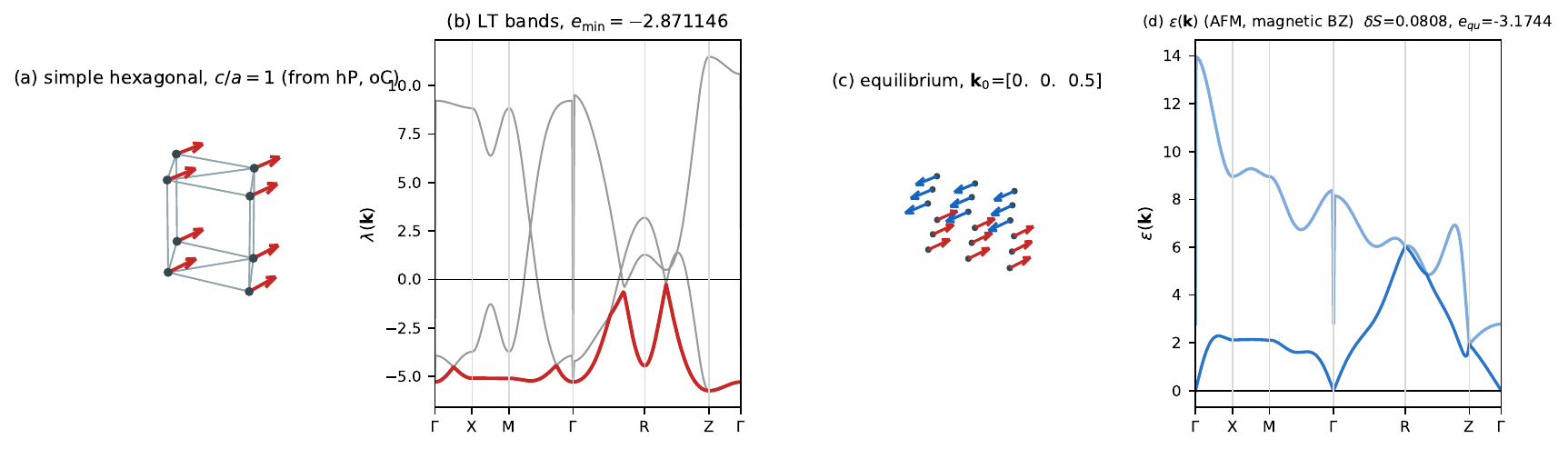}
\caption{The simple-hexagonal attractor: simple hexagonal, $c/a=1$ (from hP, oC). $e_{\min}=-2.871146$, $\mathbf k_0=(0.00,0.00,0.50)$. Panels as
in Section~\ref{sec:panels}.}
\label{fig:at3}
\end{figure*}

Figure~\ref{fig:attractors} shows the flow. Because only four distinct structures survive,
four panels characterise the optimised landscape completely, and we give them in
Figs.~\ref{fig:at2} and~\ref{fig:at3} in the same format as the representative geometries; the other two attractors coincide with representative geometries already shown (bct with Fig.~\ref{fig:p9}, simple cubic with Fig.~\ref{fig:p12}) and are not repeated.
Two of the four --- body-centred tetragonal and simple cubic --- already appear among the
representative lattices; the other two do not, and the rhombohedral attractor at
$\alpha=62.42^\circ$ is a structure that no fixed-parameter survey would have produced.

For each family the magnetic ground state is determined exactly
within the single-$\bk$ Luttinger--Tisza treatment at the optimal geometry; optimisation
over the full five-parameter metric space, without symmetry constraint, is the separate
variational problem whose answer is bct.

\subsection{The panel format}\label{sec:panels}

Figures~\ref{fig:p1}--\ref{fig:p14} present each lattice in the common four-panel format:
\textbf{(a)} the primitive cell with the equilibrium moments; \textbf{(b)} the three
Luttinger--Tisza bands $\lambda(\bk)$ along a Brillouin-zone path, the lowest band
highlighted, with $e_{\min}$; \textbf{(c)} the equilibrium configuration over several cells,
moments coloured by the sign of $\cos(\bk_0\cdot\bR)$ so that stripe and antiferromagnetic
registry are directly visible; \textbf{(d)} the magnon dispersion $\varepsilon(\bk)$ along
the same path, with $\delta S$ and $e_{\mathrm{qu}}$; the discontinuity at $\Gamma$ is
physical and is explained in Remark~\ref{rem:gammajump}. In panels (a) and (c) the three
axes carry equal data scales and the arrow length is a fixed fraction of the plotted
extent, so moments are directly comparable between figures. Panel (d) is now populated for all
thirteen collinear lattices: for the six ferromagnets it shows the single branch of
Eq.~\eqref{eq:magnon} over the crystallographic zone, and for the seven commensurate
antiferromagnets the two branches of the multi-sublattice treatment of
Section~\ref{sec:afmsw}, plotted over the \emph{magnetic} Brillouin zone of the doubled
cell. Face-centred orthorhombic carries a note in place of a curve, for the reason set out in
Section~\ref{sec:oF}.

The path (Figure~\ref{fig:bz}) runs through the corners of the reciprocal parallelepiped,
$\Gamma\to X\to M\to\Gamma\to R\to Z\to\Gamma$ in fractional coordinates
$(0,0,0)$, $(\tfrac12,0,0)$, $(\tfrac12,\tfrac12,0)$, $(\tfrac12,\tfrac12,\tfrac12)$,
$(0,0,\tfrac12)$. A generic triclinic cell has no standard high-symmetry labelling, so this
choice is used uniformly rather than adopting per-lattice crystallographic conventions;
for the high-symmetry members it coincides with the usual path.

\clearpage

\subsection{Relation to the two-dimensional programme}

The present survey completes in three dimensions what was previously done for the planar
Archimedean and Laves lattices, and the two halves differ in ways worth setting side by
side.

\begin{table*}[t]
\centering
\begin{ruledtabular}
\begin{tabular}{lll}
 & two dimensions (Archimedean $+$ Laves) & three dimensions (Bravais) \\
\colrule
family surveyed                                  & Archimedean $+$ Laves lattices & the fourteen Bravais lattices\\
$\operatorname{tr}(I-3\hat\bR\hat\bR^{T})$        & $-1$                                & $0$ identically (Proposition~\ref{prop:trace})\\
scalar shadow of the problem                     & Epstein zeta $Z_L(3/2)$, finite      & absent: $\sum_{\bR\neq0}|\bR|^{-3}$ diverges\\
behaviour at symmetric points                    & isotropic, $\Am=\alpha I$, $\alpha\neq0$ & cubic lattices: $\Am(0)\equiv0$ (Proposition~\ref{prop:cubic})\\
easy axis at the symmetric points                & selected                            & degenerate; fixed only at order $1/S$\\
rule governing the ordering type                 & lattice by lattice                  & primitive vs.\ centred (Sec.~\ref{sec:concl})\\
Luttinger--Tisza exactness                       & holds throughout                    & fails for exactly one lattice (oF, Sec.~\ref{sec:oF})\\
is the densest lattice the optimal one?          & yes (triangular)                    & \textbf{no}: fcc is densest, bct is optimal\\
optimum known in closed form                     & $-\tfrac32\zeta(3/2)L_{-3}(3/2)$    & none known\\
effect of optimising the metrics                 & not applicable (fixed geometries)   & fourteen families $\to$ four attractors\\
\end{tabular}
\end{ruledtabular}
\caption{The planar and three-dimensional dipolar surveys side by side. The last four rows
carry the physics: the Luttinger--Tisza construction is exact throughout the plane but fails
once in three dimensions; and whereas in two dimensions the energetic optimum coincides with
the densest packing --- the coincidence that links the planar problem to the classical
lattice-energy and universal-optimality literature --- in three dimensions it does not.}
\label{tab:2dvs3d}
\end{table*}

Table~\ref{tab:2dvs3d} sets the two out side by side. The last rows carry the physics. In the plane the energetic optimum coincides with the
densest packing, which is what connects the planar problem to the classical
lattice-energy and universal-optimality literature. In three dimensions that coincidence
fails: fcc is the densest Bravais packing and is beaten by bct, which forms head-to-tail
chains at lower density. The dipolar problem in three dimensions is therefore not a packing
problem in disguise, and the planar agreement is revealed as a coincidence of two dimensions.

\section{Discussion}\label{sec:discussion}

The results above are a catalogue only in their arrangement. Taken together they say
something reasonably coherent about how dipolar order is selected, and we set that out here.

\subsection{What controls the ordering wavevector}

The empirical rule --- primitive lattices order at finite $\bk$, centred lattices
ferromagnetically --- was stated by Luttinger and Tisza for the cubic
cases~\cite{LuttingerTisza1946}, and Table~\ref{tab:main} shows it holding across the whole
family. We can offer a geometric reading of why this should be so. The argument concerns the
nearest-neighbour shell, and the dipolar interaction is long ranged, so it is an
interpretation rather than a derivation; it does, however, account for the observed pattern
and indicates where the rule should be expected to weaken.

A dipolar pair is bound when it is head to tail and repelled when side by side; the crossover
is the magic angle $\arccos(1/\sqrt3)=54.74^\circ$. In a primitive lattice every site sits at
the corner of the cell, so the nearest neighbours lie along the cell edges, and a uniform
moment direction necessarily places some of those bonds beyond the magic angle. The system
recovers by reversing alternate rows: a finite-$\bk$ state converts the unfavourable bonds
into favourable ones at the cost of the favourable ones, and for a primitive lattice that
trade is profitable. Centring changes the geometry qualitatively. A body- or face-centred site
sits at the middle of the cell, so its nearest neighbours lie along body or face diagonals,
and there exists a common axis making a favourable angle with all of them simultaneously.
Uniform order is then already optimal and there is nothing for a modulation to repair.

This is why the rule is robust well outside cubic symmetry, and it also predicts its
limitations. It is a statement about which bonds dominate, so it should weaken whenever the
nearest-neighbour shell is nearly degenerate with the next --- exactly the situation at the
oF geometry discussed in Sec.~\ref{sec:oF}, which is the one lattice where the
single-$\bk$ description fails altogether.

\begin{figure}[t]
\centering
\includegraphics[width=\columnwidth]{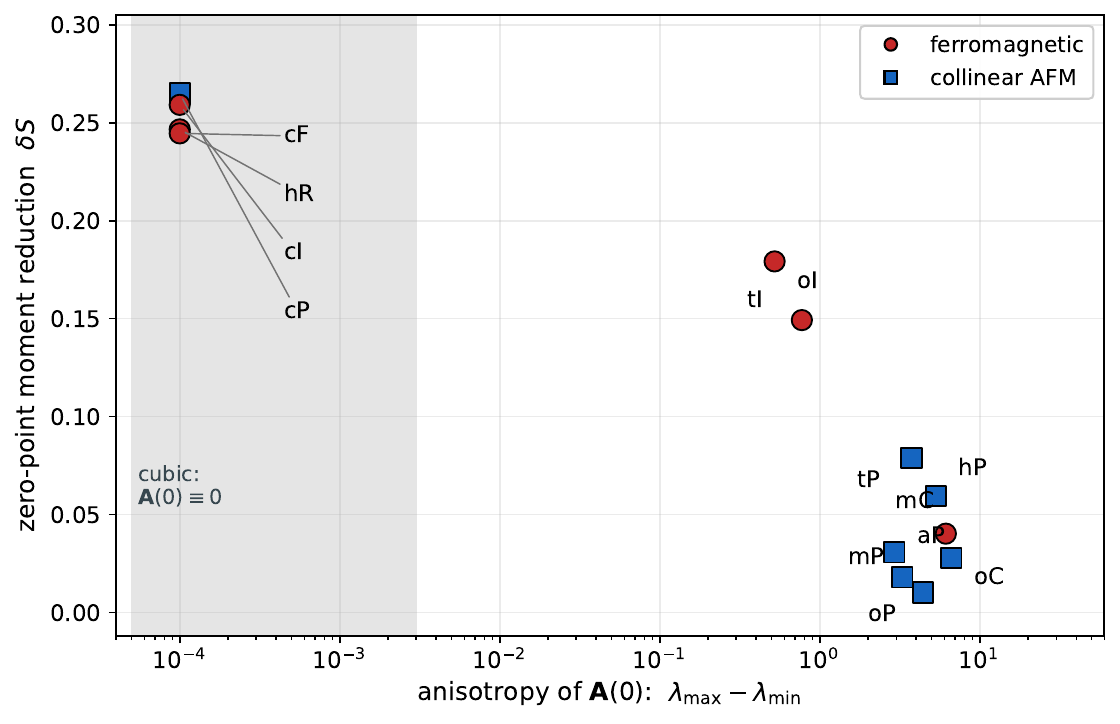}
\caption{Zero-point moment reduction against the anisotropy of the $\bk=0$ interaction
tensor, $\lambda_{\max}-\lambda_{\min}$, for the thirteen lattices with a collinear ground
state. Circles are ferromagnets, squares collinear antiferromagnets; the shaded band marks the
cubic-symmetric lattices, for which the anisotropy vanishes identically and the plotted values
are numerical zeros. The two symbol types are interleaved throughout, whereas the trend with
anisotropy is monotone across five decades: $\delta S$ measures degeneracy, not ordering type.}
\label{fig:dSaniso}
\end{figure}

\subsection{Why the zero-point reduction measures anisotropy}

The $\delta S$ column of Table~\ref{tab:main} does not sort the lattices into ferromagnets and
antiferromagnets. Empirically it sorts them by anisotropy. We offer what follows as the
physical reading of that correlation rather than as a derivation of it. The four cubic-symmetric entries have
$\delta S=0.245$--$0.265$; the weakly anisotropic bct, oI and oF follow at $0.149$--$0.179$;
the low-symmetry lattices come last at $0.010$--$0.079$, the range familiar from the planar
survey.

Figure~\ref{fig:dSaniso} makes the correlation explicit. Plotted against the anisotropy of $\Am(0)$, the thirteen collinear lattices fall on a single monotone trend spanning five decades, with ferromagnets and antiferromagnets interleaved rather than separated. The mechanism is the vanishing established in Sec.~\ref{sec:principles}. For a cubic-symmetric
lattice $\Am(0)\equiv0$, so no direction is energetically preferred at harmonic order: the
ordered moment can rotate rigidly at no cost, and the magnon spectrum acquires soft modes
throughout the zone rather than at isolated points. Since
$\delta S=\langle A_\bk/2\varepsilon_\bk-\tfrac12\rangle$ is dominated by the small-$\varepsilon$
region, a soft manifold inflates it. Anisotropy does the opposite: it gaps the transverse
fluctuations and suppresses $\delta S$.

The consequence is that $\delta S$ can be read as a diagnostic. A lattice with a large
zero-point reduction is one whose classical ground state is nearly degenerate, and in the
cubic case the degeneracy is exact at this order --- the observed easy directions of the
literature~\cite{Syro2006} being a $1/S$ effect. This also warns against interpreting the
cubic entries too literally: the axis used to compute them is an arbitrary representative of a
degenerate manifold, and only the magnitude of $\delta S$, not the direction, carries meaning.

\begin{figure}[t]
\centering
\includegraphics[width=\columnwidth]{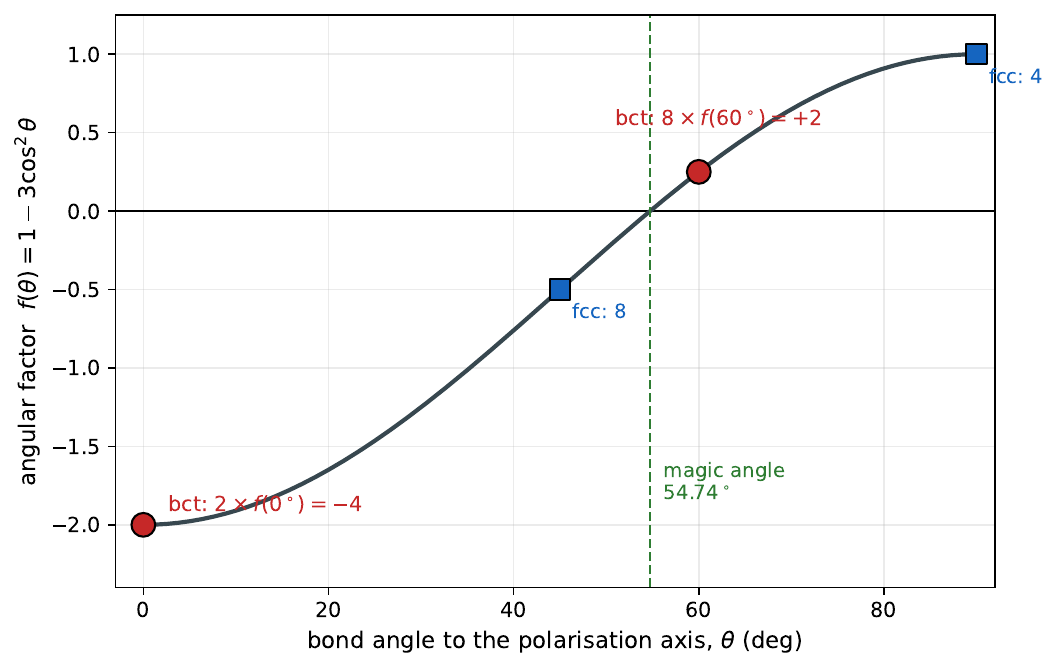}
\caption{Why the looser structure binds more strongly. The angular factor
$f(\theta)=1-3\cos^{2}\theta$ changes sign at the magic angle $54.74^\circ$. The bct shell
(circles) places two neighbours on the polarisation axis, contributing $2f(0^\circ)=-4$, and
eight at $60^\circ$, just past the sign change, costing only $8f(60^\circ)=+2$. The fcc shell
(squares) is spread too isotropically: eight neighbours at $45^\circ$ give $-4$ and four at
$90^\circ$ give $+4$, cancelling exactly, which is the vanishing of $\Am(0)$ for cubic
symmetry seen geometrically.}
\label{fig:bctfcc}
\end{figure}

\subsection{Why bct rather than fcc}

The most immediate question a reader will have is why the densest packing loses. At unit
nearest-neighbour distance fcc has $V=0.707$ against bct's $0.750$, so bct is the less dense
structure, and it wins by about three per cent.

The answer is that the dipolar interaction rewards \emph{chains}, not contacts. An isolated
head-to-tail chain of unit spacing already achieves $-2\zeta(3)=-2.404114$, which is $79\%$
of the bct value: most of the binding is intra-chain, and the lattice problem is really the
problem of packing chains without paying for it. The bct geometry at $c/a=\sqrt{2/3}$ solves
that packing problem exactly. Its coordination shell splits as $2+8$: two neighbours directly
along the moment axis, contributing the full $f(0^\circ)=-2$, and eight at precisely
$60^\circ$, where $f=+1/4$. Since the equatorial shell sits only $5.3^\circ$ beyond the magic
angle, its repulsion is eight times weaker per bond than the axial attraction, and the whole
shell costs only half of what a single chain bond gains.

Figure~\ref{fig:bctfcc} shows the contrast directly. Face-centred cubic cannot do this. Its twelve nearest neighbours split as eight at $45^\circ$ and four at $90^\circ$ about any $\langle110\rangle$ axis, contributing $8f(45^\circ)=-4$ and $4f(90^\circ)=+4$: the two cancel exactly. No axis leaves the transverse bonds near the magic angle, and by
Sec.~\ref{sec:principles} its $\bk=0$ tensor vanishes identically, so its ferromagnetic energy
is entirely the macroscopic depolarisation term. Density is simply the wrong figure of merit:
what matters is how many neighbours can be brought close to the polarisation axis, and a
slightly looser tetragonal cell does that better than the densest cubic one. This is the
sharpest contrast with the planar problem, where the optimum and the densest packing coincide.

\subsection{The rhombohedral interior optimum}

Of the families with free metric parameters, all but one flow to a boundary of their own
domain. The rhombohedral family is the exception: its optimum sits at $\alpha=62.42^\circ$,
strictly inside, with $e=-2.978125$, below the face-centred cubic value $-2.961922$ attained
at $\alpha=60^\circ$.

The reason is that the rhombohedral family passes through fcc at $\alpha=60^\circ$ and
through simple cubic at $\alpha=90^\circ$, and near fcc the cubic degeneracy is being lifted.
Increasing $\alpha$ slightly above $60^\circ$ elongates the cell along the body diagonal,
which is the polarisation axis; this lengthens the six equatorial bonds a little and shortens
nothing that matters, so the repulsive contribution falls faster than the attractive one. The
gain is small --- $0.016$, half a per cent --- and it is exhausted by $\alpha\simeq63^\circ$,
beyond which the axial bonds have stretched too far. The optimum is therefore a genuine
compromise between two competing first-order effects, which is precisely the situation that
produces an interior extremum rather than a boundary one.

We have not found a previous report of this structure in the dipolar literature, and it is a natural
candidate for experimental realisation in systems where the rhombohedral angle is tunable,
such as strained rare-earth compounds or lithographically defined arrays.

\subsection{What the four attractors mean}

Optimising within each family collapses fourteen starting points onto four end points. That
number is small for a structural reason: the families are nested, and optimisation without a
symmetry constraint is free to leave the interior of a family and terminate on its boundary,
where the symmetry is higher. What Fig.~\ref{fig:attractors} records is
therefore not fourteen independent answers but the outcome of optimising within each family
separately. This is not a proven statement about the global topology of the metric space: an
unconstrained descent from an arbitrary starting point need not follow the same route, and we
have verified the flow only along the family-wise paths.

The reading is that dipolar order in three dimensions has very few stable endpoints. Three
families reach bct, which is the global optimum; two reach simple hexagonal; two reach simple
cubic; one has its own interior minimum. A structure prepared in any of the eleven optimisable
families and allowed to relax its metric will end at one of these four, and the energies span
only $0.37$, about twelve per cent. That is a narrow window, and it suggests that in a real
material the selection between these structures will be made by whatever non-dipolar terms are
present --- exchange, strain, steric packing --- rather than by the dipolar energy itself,
which merely sets the shortlist.

\subsection{Limitations}

Three restrictions should be kept in view. The treatment is classical and single-$\bk$, which
Sec.~\ref{sec:oF} shows is not always adequate. The zero-point quantities are harmonic, so
they cannot resolve the cubic easy axis, which is a $1/S$ effect. And the survey is of Bravais
lattices: one dipole per primitive cell.

The last is the most consequential. Every crystal is a Bravais lattice decorated by a basis,
and introducing one changes the problem qualitatively rather than quantitatively. Optical
magnon branches appear; frustration can arise within the basis itself; and the structural
principles established here need not survive, since a basis site is free to sit where no
lattice point could. Whether the four-attractor picture is robust to decoration is, in our
view, the natural next question.

\section{Conclusions}\label{sec:concl}

We have determined the dipolar ground state of every three-dimensional Bravais lattice within
a single framework --- exactly, within the single-$\bk$ Luttinger--Tisza description, for
thirteen of the fourteen, and by direct supercell minimisation for the exceptional case ---
computed the linear spin-wave spectrum and zero-point corrections wherever the order is
collinear, and optimised each family over its free metric parameters. Three published values
are reproduced by methods sharing no machinery with those that produced them.

Three features of the resulting picture were not anticipated. The densest packing loses:
face-centred cubic is the densest Bravais lattice at fixed nearest-neighbour distance and is
beaten by the looser body-centred tetragonal structure, because the dipolar interaction pays
for chains rather than for contacts. In two dimensions the optimum and the densest packing
coincide, and it would have been natural to expect the same here. The zero-point reduction
sorts the lattices by degeneracy rather than by ordering type, so that the most symmetric
lattices --- the cubic ones, whose $\bk=0$ tensor vanishes identically --- carry the largest
quantum corrections. And the family-wise metric optimisation collapses fourteen starting
points onto four, with body-centred tetragonal at $c/a=\sqrt{2/3}$ the global optimum.

The robust part of the picture is the classical energetics, which rests on an Ewald engine
validated against three independent published values and internally to $\sim10^{-15}$. The
principal limitations are the restriction to one dipole per primitive cell and the harmonic,
classical treatment of the fluctuations; the natural continuation is the introduction of a
basis, where optical branches and frustration within the basis have no counterpart in the
present problem.

We would end on the methodological point, because it is the one with consequences beyond this
survey. The Luttinger--Tisza method is used routinely, and its standard caveat --- that the
relaxed solution may violate the unit-length constraint --- is usually treated as a formality.
Here it is a formality thirteen times and not the fourteenth. That the failure occurs
precisely at the one lattice with an incommensurate ordering vector, that it can be detected
by a simple test on the minimising eigenvector, and that the resulting error is not small ---
the tabulated bound lies $0.019$ below the true energy --- together suggest that the check is
worth performing routinely whenever the method is applied to a family of structures rather
than to a single one.

\begin{acknowledgments}
J.B. thanks J. Rossell\'o, Maria del Mar-, Regina-, Margalida-Batle and Maria
Vallespir-Socias for fruitful discussions. The authors received no funding for the present
research.
\end{acknowledgments}

\section*{Data Availability}
Data will be made available on reasonable request.

\end{document}